\documentclass[11pt]{article}

\usepackage{amsmath,mathtools,bbm,amssymb}
\usepackage{todonotes}
\usepackage{graphicx}
\usepackage{subcaption}
\usepackage{algorithm}
\usepackage{algpseudocode}
\usepackage{setspace}
\usepackage{natbib}
\usepackage[margin=1in]{geometry}
\usepackage[dvipsnames]{xcolor}

\makeatletter
\renewcommand*\env@matrix[1][*\c@MaxMatrixCols c]{%
  \hskip -\arraycolsep
  \let\@ifnextchar\new@ifnextchar
  \array{#1}}
\makeatother

\newcommand{\imagine}{\pmb{i}}

\newcommand{\bB}{\mathbf{B}}
\newcommand{\bD}{\mathbf{D}}
\newcommand{\bH}{\mathbf{H}}
\newcommand{\bW}{\mathbf{W}}
\newcommand{\bbeta}{\pmb{\beta}}
\newcommand{\bpsi}{\pmb{\psi}}
\newcommand{\bLambda}{\pmb{\Lambda}}
\newcommand{\bSigma}{\pmb{\Sigma}}
\newcommand{\bPsi}{\pmb{\Psi}}
\newcommand{\calM}{\mathcal{M}}
\newcommand{\calN}{\mathcal{N}}
\newcommand{\ind}{\mathbbm{1}}
\newcommand{\phe}{\phantom{{}={}}}
\newcommand{\Cov}{\mathrm{Cov}}
\newcommand{\Var}{\mathrm{Var}}
\newcommand{\Ex}{\mathbb{E}}
\DeclareMathOperator*{\argmax}{arg\,max}

\makeatletter
\DeclarePairedDelimiter\norm{\lVert}{\rVert}%
\DeclarePairedDelimiter\floor{\lfloor}{\rfloor}
\let\oldnorm\norm
\def\norm{\@ifstar{\oldnorm}{\oldnorm*}}
\makeatother

\algdef{SE}[SUBALG]{Indent}{EndIndent}{}{\algorithmicend\ }%
\algtext*{Indent}
\algtext*{EndIndent}

\title{\textbf{Hierarchical Bayesian spectral analysis of multiple stationary time series}}

\author{
\parbox{\linewidth}{\centering
Rebecca Lee$^{1}$, 
Alexander Coulter$^{1}$, Greg J. Siegle$^{2}$,\\Scott A. Bruce$^{1,*}$, and Anirban Bhattacharya$^{1}$\\$^*$sabruce@tamu.edu\\\vspace{1em}\raggedright
$^{1}$Department of Statistics, Texas A\&M University, College Station, Texas, U.S.A. \\
$^{2}$Department of Psychiatry, University of Pittsburgh, Pittsburgh, Pennsylvania, U.S.A.}}

\begin{document}

\doublespacing

\maketitle

\begin{abstract}
The power spectrum of biomedical time series provides important indirect measurements of physiological processes underlying health and biological functions.  However, simultaneously characterizing power spectra for multiple time series remains challenging due to extra spectral variability and varying time series lengths. We propose a method for hierarchical Bayesian estimation of stationary time series (HBEST) that provides an interpretable framework for efficiently modeling multiple power spectra. HBEST models log power spectra using a truncated cosine basis expansion with a novel global-local coefficient decomposition, enabling simultaneous estimation of population-level and individual-level power spectra and accommodating time series of varying lengths. The fully Bayesian framework provides shrinkage priors for regularized estimation and efficient information sharing. Simulations demonstrate HBEST's advantages over competing methods in computational efficiency and estimation accuracy. An application to heart rate variability time series demonstrates HBEST's ability to accurately characterize power spectra and capture associations with traditional cardiovascular risk factors.
\end{abstract}


\section{Introduction}
\label{s:intro}

The frequency domain characteristics of biomedical time-series data, such as heart rate variability, electroencephalography, and functional magnetic resonance imaging, provide important indirect measurements of underlying physiological processes associated with health and functioning \citep{hall2004,knyazev2012,yuen2019intrinsic}.  To better understand health and functioning on a broader scale, biomedical studies typically collect and analyze time series data from multiple participants representing a specific population of interest.  This allows researchers to develop both population-level summary measures associated with health and functioning, as well as variability in summary measures within the population.  Frequency domain properties of time series data can be characterized through the power spectrum, which has a direct physical interpretation \citep{priestley1981} and is a natural starting point for formulating models \citep{diggle_spectral_1997}. 

A prime example comes from our motivating longitudinal study on cardiovascular disease risk factors in a diverse population of older adults \citep{zhang_national_2018,chen_racialethnic_2015}.  The Multi-Ethnic Study of Atherosclerosis (MESA) recruited a diverse population-based sample of participants age 45 and older to study subclinical cardiovascular disease progression and predictive markers associated with disease progression \citep{bild2002mesa}. A subset of 2,060 participants also completed a single-night unattended in-home sleep study in which heart rate variability (HRV), among other signals, was collected to better understand the association between sleep characteristics and cardiovascular risk factors.  HRV is a measure of the elapsed time between consecutive heart beats, and its power spectrum provides objective measures of stress and arousal \citep{hall2004}.  Lower HRV high-frequency power is also associated with an increased risk of cardiovascular disease \citep{hillebrandetal2013}.  

\begin{figure}[ht!]
    \centering
    \begin{subfigure}[b]{0.9\textwidth}
        \centering
        \includegraphics[trim={0 .7cm 0 0},clip,width=.90\linewidth]{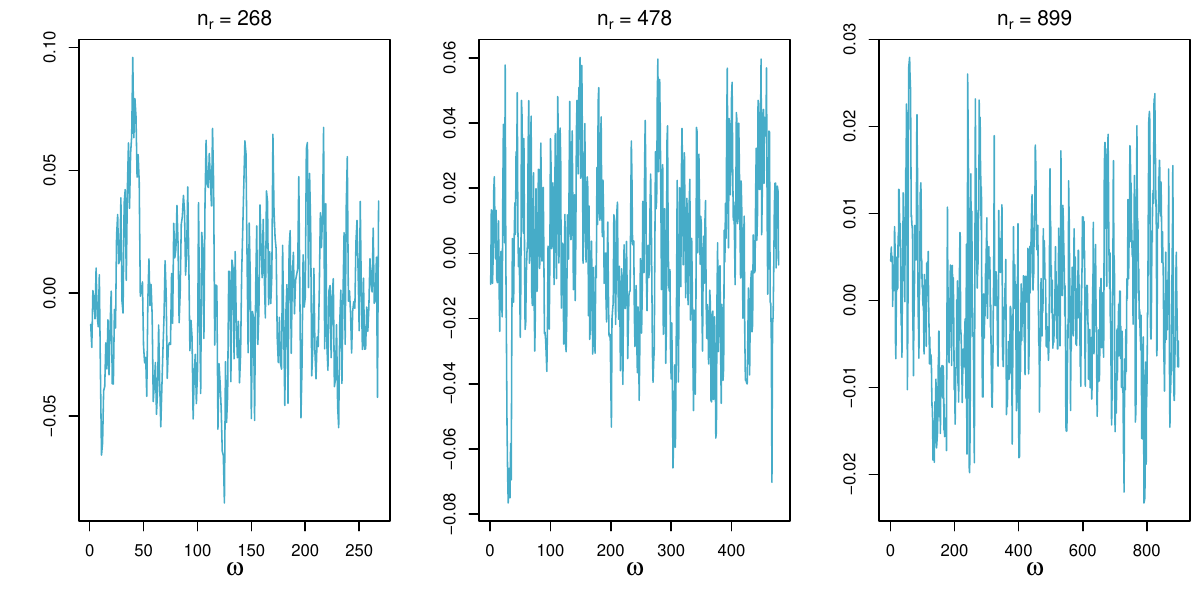}
        \caption{}
        \label{fig:MESA_timeseries}
    \end{subfigure}
    \begin{subfigure}[b]{0.9\textwidth}
        \centering
        \includegraphics[width=.90\linewidth]{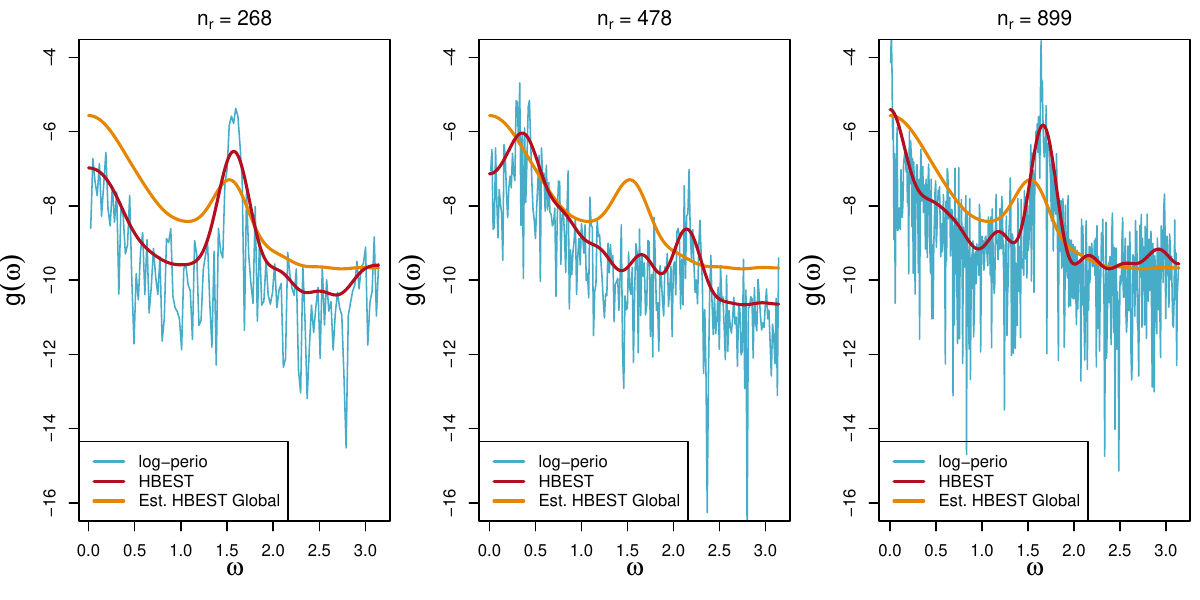}
        \caption{}
        \label{fig:MESA_logspectra}
    \end{subfigure}
    
    \caption{(a) Standardized heart rate variability (HRV) time series with 1 Hz sampling rate during the first onset of non-rapid eye movement (NREM) sleep during a single night, unattended, in home sleep study for three participants from the MESA study.  These participants represent the 25th, 50th, and 75th percentiles (\emph{left to right}) of time series length. (b) Estimates of the HRV log power spectrum ($g(\omega)$) are provided for each time series including the log-periodogram (blue), population-level estimate of the log power spectrum from the proposed HBEST method (orange), and individual-level estimate of the log power spectrum from the proposed HBEST method (red).}
    \label{fig:MESA}
\end{figure}

Figure \ref{fig:MESA} displays HRV time series and estimates of the power spectrum during the first onset of non-rapid eye movement (NREM) sleep in three MESA participants.  Our analysis seeks to provide accurate estimation of both population-level and individual-level power spectra by sharing information about similarities in the shape of the power spectra across time series, improving accuracy for shorter time series in particular, since they do not carry as much independent information about the underlying power spectra.  Additionally, we seek to provide appropriate uncertainty quantification for estimates of the power spectra by properly characterizing variability within the population.  

Estimation of the power spectrum for a \textit{single} time series has been well studied; see \cite{shumway_time_2017} for an overview of various parametric and nonparametric approaches.  In particular, basis expansion methods have been widely used for nonparametric estimation of the power spectrum, using various bases including Bernstein polynomials \citep{choudhuri_bayesian_2004}, second-order autoregressive kernels \citep{granados-garcia_brain_2022}, intrinsic mode functions \citep{Huang1998}, and cosine functions \citep{rosen_local_2009, rosen_adaptspec_2012}.  

However, accurate estimation of power spectra from \textit{multiple} time series is more challenging. Power spectra typically exhibit similarities across the multiple time series, but they also carry extra spectral variability and cannot be assumed to share a single common power spectrum \citep{diggle_spectral_1997, iannaccone2001, krafty_functional_2011}. Failing to account for this extra variability will lead to poor uncertainty quantification at the population level, biased estimation at the individual level, and overall faulty inference. On the other hand, modeling each power spectrum independently, thus ignoring potential similarities in their shapes, precludes the sharing of information and results in less efficient estimation \citep{hart_nonparametric_2022}. Second, time series in practice often have different lengths, leading to different frequency grids over which the estimates of the power spectrum are typically computed.  This makes it difficult to share information when estimating individual power spectra and to estimate the population-average power spectrum \citep{Caiado2009}.

Although several recent methods have been developed, there remain important challenges and opportunities for improvement. \cite{cadonna_bayesian_2019} proposes a Gaussian mixture modeling approach to approximate the likelihood for a collection of power spectra, which allows sharing of information to improve individual estimates.  However, this method does not model the population-level power spectra explicitly.  \cite{hart_nonparametric_2022} proposes a method using a Bernstein polynomial basis representation with an additional grouping structure to inform individual estimates of the power spectra.  However, this approach doesn't allow for variation in the power spectra within groups.  

The proposed method provides an interpretable nonparametric framework for modeling power spectra of multiple time series drawn from a population of interest. Our Hierarchical Bayesian Estimation of Stationary Time series (HBEST) approach allows for sharing of information across replicates and provides direct estimates of both population-level and individual-level power spectra with appropriate uncertainty quantification for both.  Using a Demmler-Reinsch basis expansion, HBEST represents the log power spectra as a linear combination of cosine basis functions and further decomposes the coefficients for each basis function as the sum of two parts: a population-level component and an individual-level component for improved interpretability.  Priors are imposed on model parameters in a fully Bayesian hierarchical model that encourages regularized estimation of power spectra for consistent estimation, and sharing of information across power spectra regarding the shape of the spectra when appropriate for improved estimation accuracy.  

The proposed method makes important contributions to the literature on spectral analysis of multiple time series in three important ways.  First, by explicitly modeling both population- and individual-level power spectra, the model offers improved interpretability and estimation at both levels.  Second, the model allows for extra spectral variability commonly encountered in practice \citep{diggle_spectral_1997}, which is crucial for accurate uncertainty quantification.  Lastly, the truncated cosine basis representation for the log power spectra used in this work provides a scalable and computationally efficient alternative to other basis representations (e.g. Bernstein polynomials), which can require significantly more basis functions to achieve comparable accuracy \citep{Farouki2012Bernstein}, leading to increased computational complexity and slower run times.

\textbf{Notation.} We denote scalars as $a, A, \alpha \in \Re$; vectors as $\mathbf{d}, \bbeta \in \Re^p$; and matrices as $\mathbf{B}, \pmb{\Sigma} \in \Re^{n\times p}$. For consistency, we exclusively reserve $\ell$ to index time series $\ell = 1, \hdots, L$, and exclusively reserve $b$ to index Demmler-Reinsch basis functions and coefficients $b = 1, \hdots, B$. We similarly use condensed notation $\{d_b\} \equiv \{d_b\}_{b = 1}^B$, $\{\zeta_{\ell}\} \equiv \{\zeta_{\ell}\}_{\ell=1}^L$, and $\{\beta_{\ell, b}\} \equiv \{\beta_{\ell, b}\}_{\ell=1,b=1}^{L, B}$ to refer to collections of values aggregating over the given index sets. We denote the indicator function $\ind\{\:\cdot\:\}$ -- that is, $\ind\{\mathrm{true}\} = 1$, and zero otherwise -- and we denote the imaginary unit $\imagine$.

\section{Modeling Framework}\label{s:model}

\subsection{Background} Consider a collection of $L$ zero-mean {\it stationary} time series, $\{X_{\ell, t}\}$. The frequency domain characteristics of each series can be obtained from its {\it spectral representation} \citep{cramer42, priestley1981},
$
X_{\ell, t} = \int_{0}^{2\pi}A_{\ell}(\omega)\exp\{\imagine \omega t \} dZ_{\ell}(\omega),
$
where $A_{\ell}(\omega)$ is a complex-valued function of frequency $\omega$ that is $2\pi$-periodic and Hermitian $A_{\ell}(2\pi-\omega) = \overline{A_{\ell}(\omega)}$, and $dZ_{\ell}(\omega)$ is a zero-mean orthogonal process with unit variance. Provided the autocovariance function of $X_{\ell, t}$ is absolutely summable, the power spectrum of the series is defined as $f_{\ell}(\omega) = |A_{\ell}(\omega)|^2$ and can be interpreted as the contribution to the variance of $X_{\ell, t}$ from oscillations at frequency $\omega$, and is the primary tool used to describe the frequency-domain properties of stationary time series.

Given a realization of the $\ell^{\mathrm{th}}$ time series, $x_{\ell, 1},\hdots,x_{\ell n_{\ell}}$, the {\it periodogram} at frequency $\omega$ is $Y_{\ell}(\omega) = \frac{1}{n_{\ell}}\left| \sum_{t=1}^{n_{\ell}} x_{\ell, t}\exp\{ - \imagine \omega t \} \right|^2.$ While the periodogram is an unbiased estimator of the true spectrum and periodogram ordinates are asymptotically ($n_{\ell} \rightarrow \infty$) uncorrelated, the asymptotic variance does not tend to zero. As such, smoothed estimators that share information across frequencies are needed. Among many possible choices, we adopt the Bayesian penalized linear spline model \citep{wahba1980,ruppert2003} as a flexible modeling paradigm that can be seamlessly extended to more complex settings like ours. In previous work, linear splines have been found to be more accurate in capturing changes in the power spectra compared to higher order splines \citep{rosen_adaptspec_2012}.

\subsection{HBEST Model}

With this background, we now lay down our proposed hierarchical model HBEST. Recall we have a collection of $L$ many zero-mean stationary time series, $\{X_{\ell, t}\}$, with realization $\{x_{\ell, t}\}_{t=1}^{n_\ell}$ from the $\ell^{\mathrm{th}}$ series/replicate. We allow the lengths of the replicates $n_\ell$ to vary across $\ell$ in the subsequent development. Let $g_\ell(\omega) := \log f_\ell(\omega)$ denote the log-spectrum for the $\ell^{\mathrm{th}}$ replicate. As in \cite{rosen_adaptspec_2012}, we consider expanding each $g_\ell$ over a (truncated) Demmler-Reinsch basis expansion \citep{schwarz2016},
\begin{equation}\label{eq:DR_basic}
g_\ell(\omega) = \alpha_\ell + \sum_{b = 1}^B \beta_{\ell, b} \psi_b(\omega), \qquad \psi_b(\omega) := \sqrt{2}\, \cos(\omega b), \qquad \omega \in [0, 2\pi]. 
\end{equation}
Here, $\alpha_\ell$ is a replicate-specific intercept; $\bbeta_{\ell} := ( \begin{matrix} \beta_{\ell,1} & \cdots & \beta_{\ell,B} \end{matrix} )^{\top}$ is a replicate-specific vector of basis coefficients, where $B$ is the number of cosine basis functions employed (uniformly across all replicates); and $\psi_b(\:\cdot\:)$ denotes the $b^{\mathrm{th}}$ cosine basis function. In light of \eqref{eq:DR_basic}, the task of hierarchically modeling the $g_\ell$'s boils down to the same for $(\alpha_\ell, \bbeta_{\ell})$ across $\ell$. To that end, we consider an additive ANOVA-type {\it global-local} decomposition of the coefficients
\begin{align}\label{eq:anova_decomp}
\alpha_\ell = \alpha^{\mathrm{glob}} + \alpha^{\mathrm{loc}}_\ell, \qquad \beta_{\ell, b} = \beta^{\mathrm{glob}}_b + \beta^{\mathrm{loc}}_{\ell, b}.
\end{align}
The decomposition \eqref{eq:anova_decomp} expresses $\beta_{\ell, b}$, the $b^{\mathrm{th}}$ basis coefficient for the $\ell^{\mathrm{th}}$ replicate, as the sum of a {\it global} basis-specific term $\beta^{\mathrm{glob}}_b$ and a {\it local} replicate plus basis specific term $\beta^{\mathrm{loc}}_{\ell, b}$. A similar decomposition is employed for the replicate-specific intercept $\alpha_\ell$. Overall, \eqref{eq:anova_decomp} translates into an additive {\it global-local} decomposition of the replicate-specific log-spectra $g_\ell$ as 
\begin{align}\label{eq:gld}
g_\ell(\omega) = g^{\mathrm{glob}}(\omega) + g^{\mathrm{loc}}_\ell(\omega), \qquad \omega \in [0, 2\pi],
\end{align}
where $g^{\mathrm{glob}}(\omega) = \alpha^{\mathrm{glob}} + \sum_{b=1}^B \beta^{\mathrm{glob}}_b \, \psi_b(\omega)$
and $g^{\mathrm{loc}}_\ell(\omega) = \alpha^{\mathrm{loc}}_\ell + \sum_{b=1}^B \beta^{\mathrm{loc}}_{\ell, b} \, \psi_b(\omega)$. This amounts to a {\it multiplicative} decomposition $f_\ell(\omega) = f^{\mathrm{glob}}(\omega) \, f^{\mathrm{loc}}_\ell(\omega)$ for the untransformed spectra.

We now discuss priors on the constituent terms in \eqref{eq:anova_decomp}. We assume independent Gaussian priors $\alpha^{\mathrm{glob}} \sim \mathcal{N}(0, \sigma_\alpha^2)$ and $\alpha^{\mathrm{loc}}_\ell \stackrel{\mathrm{iid}}{\sim} \mathcal{N}(0, \delta^2)$. Here, $\sigma_\alpha^2$ is typically set to a large value (e.g., 100) to impose a diffuse prior on the global intercept $\alpha^{\mathrm{glob}}$, while $\delta^2 \ll 1$ is set to a small value to make the priors on the local intercept terms $\alpha^{\mathrm{loc}}_\ell$'s more tightly concentrated around zero. Next, given a global scale parameter $\tau > 0$, basis-specific scale parameters $d_b > 0$, and replicate-specific scale parameters $\xi_\ell \in (1, \infty)$, we propose {\it independent} zero-mean conditional Gaussian priors for $\{\beta^{\mathrm{glob}}_b\}$ and $\{\beta^{\mathrm{loc}}_{\ell, b}\}$ with a multiplicative decomposition of their respective variances
\begin{align}\label{eq:prior_betabl}
\beta^{\mathrm{glob}}_b \mid \tau, d_b  \stackrel{\mathrm{ind}}{\sim} \mathcal{N}(0, \tau^2 d_b), \qquad \beta^{\mathrm{loc}}_{\ell, b} \mid \tau, d_b, \zeta_\ell \stackrel{\mathrm{ind}}{\sim} N\big(0, \tau^2 d_b (\zeta_\ell^2 - 1) \big).
\end{align}
In \eqref{eq:prior_betabl}, $\{d_b\}$ is a deterministic {\it decreasing} sequence that encourages higher degree of shrinkage for higher-frequency basis functions; following \cite{li_adaptive_2019} we set $d_b := (4 \pi b^2)^{-1}$; when modeling a single stationary time series, this choice of $d_b$ leads to a $\tau^2 \chi_B^2$ distribution on the squared $L_2$ norm of the log-spectra. The global scale parameter $\tau$ controls overall smoothness of the spectra while the replicate specific local parameters $\zeta_\ell$ allow for additional variability at the individual spectra level. We endow $\tau$ with a standard Half-$t$ prior \citep{gelman2006prior,polson2012local} with degrees of freedom $\nu_{\tau}$, and place independent Half-$t$ priors (with degrees of freedom $\nu_{\zeta}$) restricted to $(1,\infty)$ on the $\zeta_\ell$'s. A salient feature of our prior construction is that it induces a similar prior on $g^{\mathrm{glob}}$ as \citet[Section 2.2]{rosen_adaptspec_2012}, while allowing replicate-specific departures via $g_\ell^{\mathrm{loc}}$ that are simultaneously {\it a priori} strongly shrunk towards zero and possess heavy tails \citep{carvalho2010horseshoe}. This {\it adaptive} shrinkage allows for {\it borrowing of information} across the replicates when warranted. This completes the HBEST specification.

It follows from \eqref{eq:prior_betabl} that under HBEST, $\beta_{\ell, b} \mid \tau, d_b, \zeta_\ell \sim \mathcal{N}(0, \tau^2 d_b \zeta_\ell^2)$ for any fixed $\ell, b$. The multiplicative decomposition of the variance therefore provides a bi-directional shrinkage on the matrix of basis coefficients $\beta_{\ell, b}$ which is exchangeable across $\ell$ and more aggressively shrunk with increasing $b$. To further investigate this, we study the row- and column-wise shrinkage profiles induced by HBEST; see Appendix A for detailed derivations.
We denote ``row" vectors $\bbeta_{\ell\bullet} := (\begin{matrix} \beta_{\ell, 1} & \cdots & \beta_{\ell, b} \end{matrix})^{\top}$ and ``column" vectors $\bbeta_{\bullet b} := ( \begin{matrix} \beta_{1, b} & \hdots & \beta_{L, b} \end{matrix} )^{\top}$. The shared dependence on $\beta_b^{\mathrm{glob}}$ induces dependence between $\beta_{\ell, b}$ and $\beta_{{\ell', b}}$ ($\ell \ne \ell'$) given the hyperparameters, leading to a non-diagonal covariance matrix of $\bbeta_{\bullet b}$. Specifically, one obtains
\begin{align*}
    \bbeta_{\ell\bullet} \mid \tau, \zeta_{\ell}, \{d_b\} &\sim \mathcal{N}_B\left( \pmb{0}, \tau^2 \zeta_{\ell}^2 \mathbf{D} \right), \\
    \bbeta_{\bullet b} \mid \tau, \{\zeta_{\ell}\}, d_b &\sim \mathcal{N}_L\left( \pmb{0}, \tau^2 d_b (\mathrm{diag}(\zeta_{\ell}^2 - 1) + \pmb{1}\pmb{1}^{\top}) \right),
\end{align*}
where $\mathbf{D} := \mathrm{diag}( \begin{matrix} d_1 & \cdots & d_B \end{matrix} )$ and $\pmb{1} \equiv \pmb{1}_L$ is an $L$-dimensional vector of ones. It follows the matrix of coefficients $\mathbf{B} \equiv (\beta_{\ell, b}) \in \Re^{L \times B}$ conditionally follows a matrix normal distribution
\begin{align*}
\mathbf{B} \mid \tau, \{\zeta_\ell\}, \{d_b\} \sim \mathcal{MN}_{L, B}\left( \pmb{0}, \mbox{diag}(\zeta_\ell^2-1) + \mathbf{1} \mathbf{1}^{\top}, \tau^2 \mathbf{D} \right). 
\end{align*}

\subsection{Posterior Computation}

For each $\ell$ and $j_{\ell} = 1, \hdots, m_{\ell} := \floor{\frac{n_{\ell}}{2}}$, let $\omega_{j_{\ell}}^\star := 2 \pi j_{\ell}/n_\ell$ denote the {\it fundamental frequencies}, and $Y_{j_{\ell}} \equiv Y_\ell(\omega_{j_{\ell}}^\star)$ denote the corresponding periodogram ordinates. We assume the $Y_{j_{\ell}}$'s are conditionally independent across $\ell$. Moreover, for each $\ell$, we use Whittle's likelihood~\citep{whittle1953} approximation to model the $Y_{j_{\ell}}$'s (across $j$) as conditionally independent exponential random variables, each with mean $f_\ell(\omega_{j_{\ell}}^\star)$. The ensuing product Whittle likelihood is combined with the proposed hierarchical prior on the model parameters. We sample from the resulting posterior distribution via a highly efficient Metropolis-within-Gibbs sampler, which we elaborate in Appendix B; specific algorithms for steps of the sampler are provided in Appendix C. Briefly, we separately block-update $\{\beta_b^{\mathrm{glob}}\}$ and $\{\beta_{\ell, b}^{\mathrm{loc}}\}$ using independence Metropolis proposals based on Gaussian approximations to their respective (joint) conditional posteriors, which are conveniently derived exploiting log-concavity of the Whittle likelihood. We update global scale parameter $\tau$ using a Griddy Gibbs sampler on a grid $[\tau_{\min}, \tau_{\max}]$, and similarly update local scale parameters $\{\zeta_\ell\}$ on a common grid $[\zeta_{\min}, \zeta_{\max}]$.

\section{Simulations}
\label{s:sims}

To assess finite-sample performance, we compare HBEST with the method of \cite{hart_nonparametric_2022}, herein referred to as the ``Hart model".  The Hart model allocates time series to groups and estimates shared within-group spectral densities using Bernstein polynomials in a fully Bayesian model. Posterior inference on the number of groups and group membership is handled with a nested Dirichlet process. The available Hart model implementation requires standardized (mean zero and unit variance) time series of equal length. While HBEST can natively handle multiple time series of different lengths without standardization, we adopt these constraints for a fair comparison.

We simulated $S=30$ sets of data from two different settings. Each set of data contained $L=15$ conditionally independent time series of equal length, $n_{\ell} \equiv n = 1000$ for all $\ell$, standardized to have zero mean and unit variance. This $n$ was chosen since it is close to the average length of the HRV time series from the MESA sleep study. We consider $5000$ MCMC iterations with a burn-in period of $500$, and collect $I = 4500$ posterior samples post burn-in. For HBEST, we use $B = 15$ basis functions. Lastly, we chose the following values for hyperparameters in the HBEST model: $\nu_{\tau} = 2$; $\sigma^2_{\alpha} = 100$; $\delta^2 = 0.1$; $\nu_{\zeta} = 5$; $\zeta_{min} = 1.001$; $\zeta_{max} = 15$; $\tau_{min} = 0.001$; $\tau_{max} = 100$. The Hart model adaptively determines the number of Bernstein polynomials used to represent approximate power spectra.  We set the maximum number of Bernstein polynomials to be 30.  

Let $K \gg 0$ and $\mathcal{K} := \{\frac{a}{K-1} : a = 0, 1, \dots, K-1\}$, and let $\omega_k = \pi k$ for $k \in \mathcal{K}$. To measure estimation accuracy, we compute the trimmed Approximate Expected Posterior Loss (AEPL) for each data set as
\begin{equation}\label{eq.MSE_IndCom}
    AEPL = \{LI|\mathcal{J}|\}^{-1}\sum_{\ell=1}^{L}\sum_{i=1}^{I}\sum_{j\in\mathcal{J}}[\hat{g}_{\ell}^{i}(\omega_{j}) - g_{\ell}(\omega_{j})]^2, 
\end{equation}
where $\mathcal{J} := \mathcal{K} \cap [0.05, 0.95)$ has cardinality $|\mathcal{J}|$, $\hat{g}_{\ell}^{i}(\omega)$ is the estimate of the log power spectrum for the $\ell^{\mathrm{th}}$ time series from the $i^{\mathrm{th}}$ iteration of the sampler, and $g_{\ell}(\omega)$ is the true log power spectrum for the $\ell^{\mathrm{th}}$ time series. We trimmed the index set $\mathcal{K}$ to $\mathcal{J}$ for a fairer comparison since the estimation accuracy using the Bernstein polynomial basis representation for frequencies near 0 and $\pi$ is quite unstable. We use $K=1000$.

\subsection{Conditional MA(4) with Multiple Levels of Variation} The first is a conditional MA(4) model used in \cite{granados-garcia_brain_2022} of the form 
$
    X_{\ell, t} = \theta_{\ell, 1}\epsilon_{\ell, t-1} + \theta_{\ell, 2}\epsilon_{\ell, t-2} + \theta_{\ell, 3}\epsilon_{\ell, t-3} + \theta_{\ell, 4}\epsilon_{\ell, t-4} + \epsilon_{\ell, t},
$
where $\epsilon_{\ell, t}$ is a white noise process with unit variance. For the first setting, we consider the case where all time series share a common underlying power spectra such that $(\theta_{\ell, 1}, \theta_{\ell, 2}, \theta_{\ell, 3}, \theta_{\ell, 4}) \equiv (\theta_1, \theta_2, \theta_3, \theta_4) = (-0.3, -0.6, -0.3, 0.6)$ for all $L$ time series. For the second and third setting, we add variation in the spectra among the time series by randomly sampling the first coefficient $\theta_{\ell, 1} \sim \mathcal{N}(-0.3, 0.09\alpha^2)$, with $\alpha=0.15$ for the moderate variation setting and $\alpha=0.3$ for the high variation setting. This allows for a more detailed comparison of performance under different levels of variability within the population.

\subsection{Conditional Mixture of AR(2) Processes} The second setting generates time series as conditional mixtures of AR(2) processes that allow for moderate variation across replicates. Following \cite{granados-garcia_brain_2022}, we generate time series as the sum of two AR(2) processes $X_{\ell, t} = Z_{1\ell, t} + Z_{2\ell, t}$
where 
$
     Z_{i\ell, t} = \phi_{i\ell, 1}Z_{i\ell, t-1} + \phi_{i\ell, 2}Z_{i\ell, t-2} + \epsilon_{i\ell, t}, \; i=1,2,
$
and $\epsilon_{i\ell, t}$ is a white noise process with unit variance. The coefficients $\phi_{i\ell, 1}= 2\cos(\gamma_{ i \ell})\exp\{-\kappa_{i\ell}\}$ and $\phi_{i\ell, 2} = -\exp\{-2\kappa_{i\ell}\}$ are parameterized with respect to the frequencies representing the localized peaks in the power spectrum ($\gamma_{i \ell }$) and the bandwidths representing the spreads ($\kappa_{i\ell}$). We generated peaks and bandwidths for the two AR(2) processes for each time series as follows: $\gamma_{1\ell} \sim U(0.2,0.23)$, $\kappa_{1\ell} \sim U(0.1,0.2)$, $\gamma_{2\ell} \sim U(\pi/5-0.1,\pi/5+0.1)$, $\kappa_{2\ell}=0.15$. See Figure \ref{fig:Sims_mixture_comparison} for the true population average log power spectrum. This particular setting was chosen since it mimics the shape and variability in the HRV log spectra seen among MESA participants.

\subsection{Results} 
\begin{figure}[ht!]
    \centering
     \begin{subfigure}[b]{0.9\textwidth}
        \centering
    \includegraphics[width=.45\linewidth]{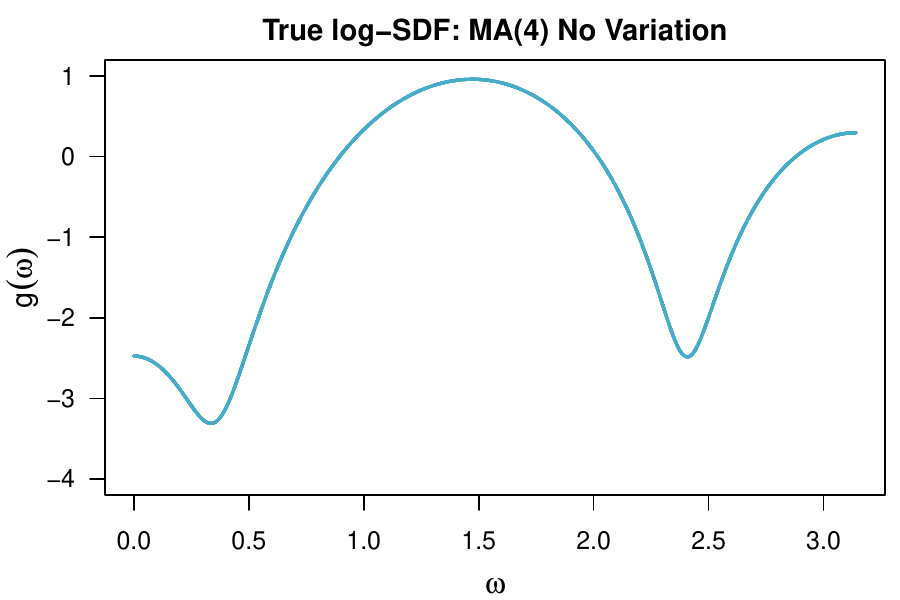}
     \includegraphics[width=.45\linewidth]{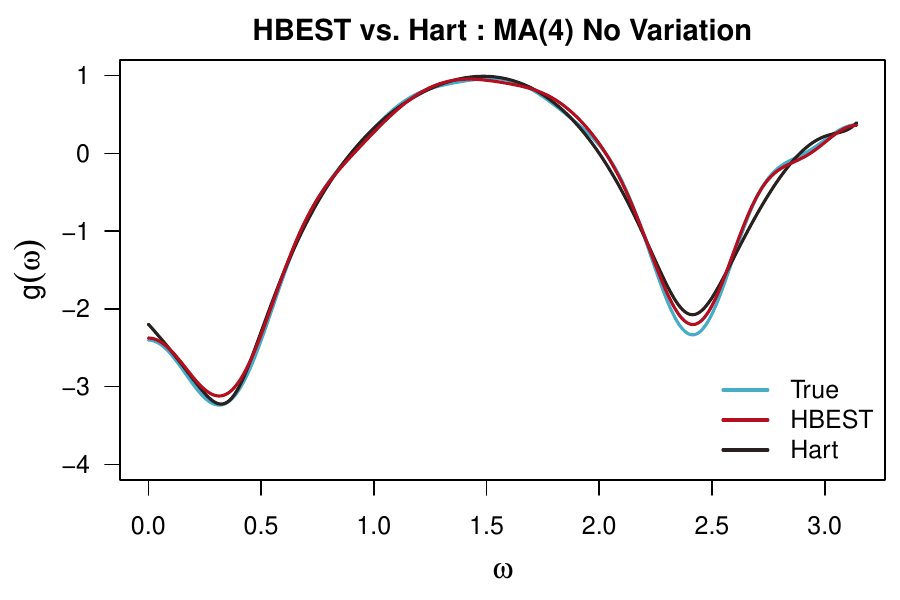}
      \caption{}
        \label{fig:Sims_fit_novar}
    \end{subfigure}
     \begin{subfigure}[b]{0.9\textwidth}
        \centering
    \includegraphics[width=.45\linewidth]{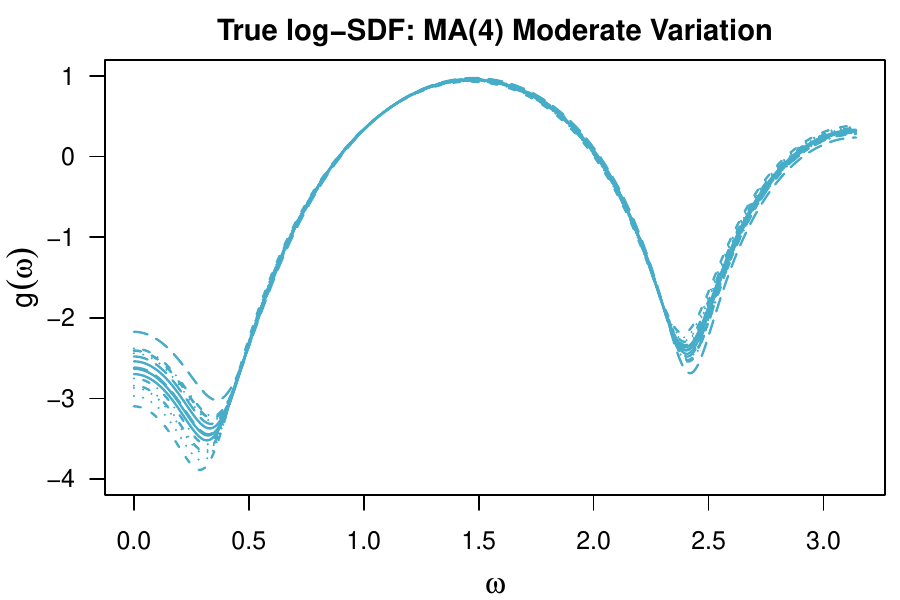}
    \includegraphics[width=.45\linewidth]{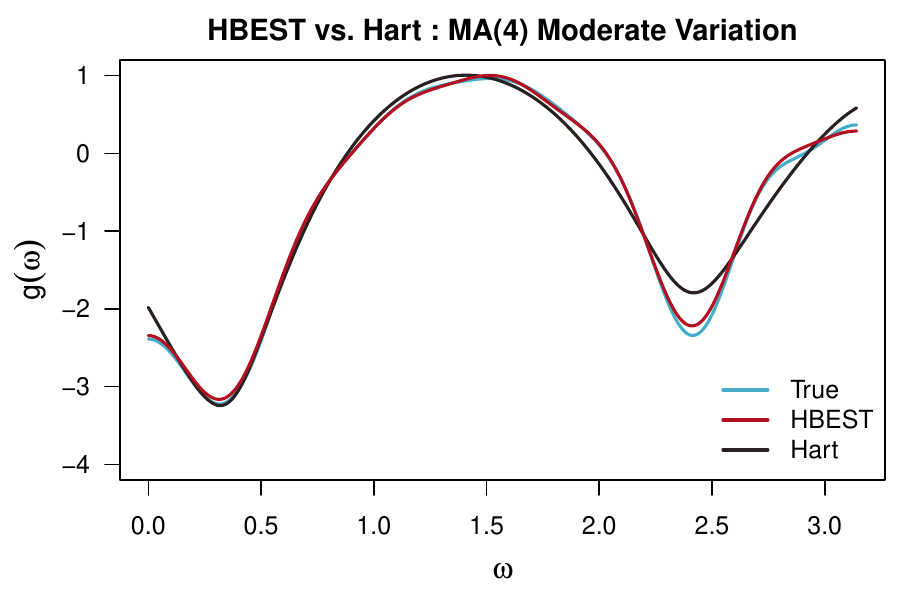}
     \caption{}
        \label{fig:Sims_fit_medvar}
    \end{subfigure}
     \begin{subfigure}[b]{0.9\textwidth}
        \centering
     \includegraphics[width=.45\linewidth]{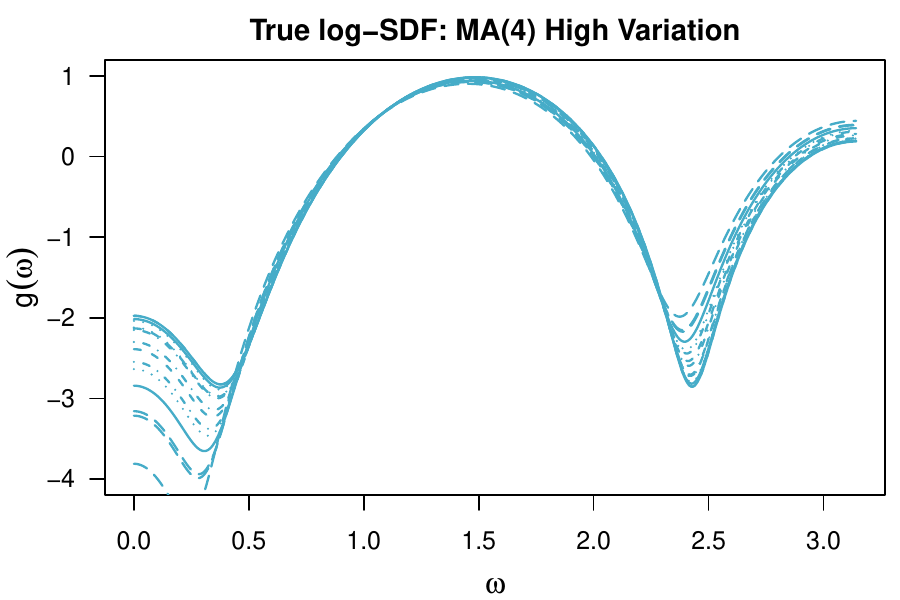}
     \includegraphics[width=.45\linewidth]{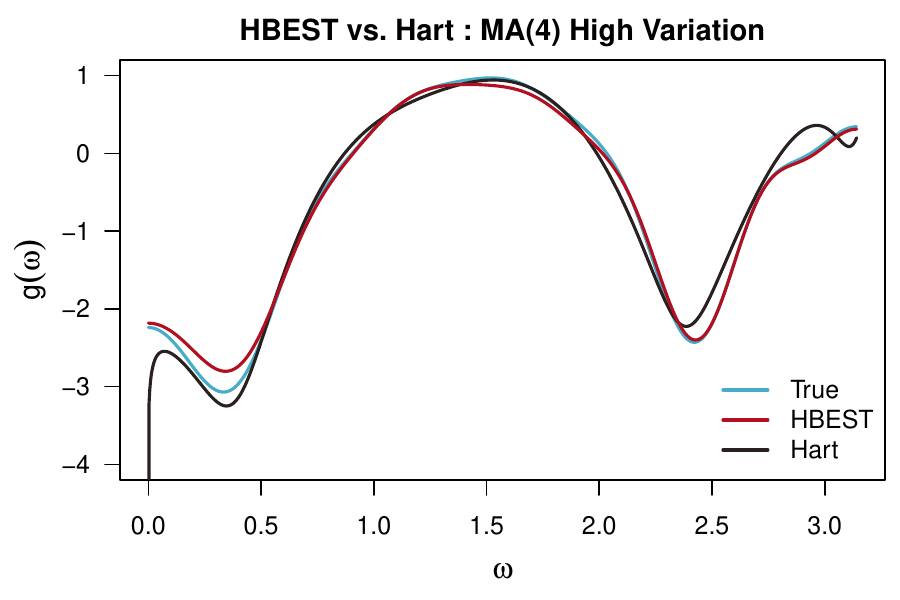}
      \caption{}
        \label{fig:Sims_fit_highvar}
    \end{subfigure}
    \caption{True individual level log power spectra for the three conditional MA(4) settings (left) and comparison of estimates for an individual log power spectrum using the proposed HBEST method and the method from \protect\cite{hart_nonparametric_2022} (right). Each row corresponds to a different level of variation in the true log power spectra: (a) no variation, (b) moderate variation, and (c) high variation.
    }
    \label{fig:Sims_fit}
\end{figure}

\begin{figure}[ht!]
    \centering
    \includegraphics[width=0.65\linewidth]{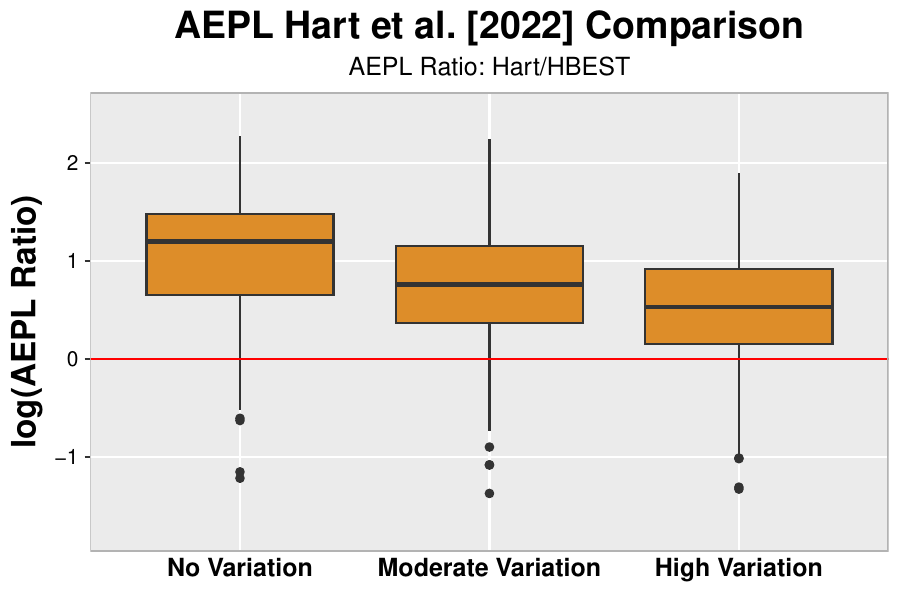}
    \caption{Distribution of log ratios of the Approximate Expected Posterior Losses (AEPLs) \eqref{eq.MSE_IndCom} across simulated datasets for HBEST and \cite{hart_nonparametric_2022} for the conditional MA(4) settings. Median and mean AEPLs for each variation setting for each model (HBEST, Hart) with lowest AEPL bolded: No variation: median: ($\pmb{0.01}$, $0.03$); mean: ($\pmb{0.01}$, $0.03$). Moderate variation: median: ($\pmb{0.01}$, $0.04$); mean: ($\pmb{0.02}$, $0.07$). High variation: median: ($\pmb{0.02}$, $0.04$); mean: ($\pmb{0.03}$, $0.05$) 
    }
    \label{fig:Sims_AEPLratio}
    \vspace{2cm}
\end{figure}

Figure \ref{fig:Sims_fit} shows examples of posterior mean estimators of the power spectrum for a single replicate for the three levels of variation from the conditional MA(4) setting. Figure \ref{fig:Sims_AEPLratio} shows the box-plot of the AEPL ratio across datasets comparing the two methods for each setting. Across all levels of variability, HBEST outperforms the Hart model as the log AEPL ratio exceeds 0 for nearly all simulated datasets.
We see that as variability within the population increases, the performance gap between HBEST and the Hart model shrinks. This is expected since the Hart model can account for some of the variability by grouping series with similar power spectra for improved estimation.  

\begin{figure}[ht!]
    \centering
    \includegraphics[width=.49\linewidth]{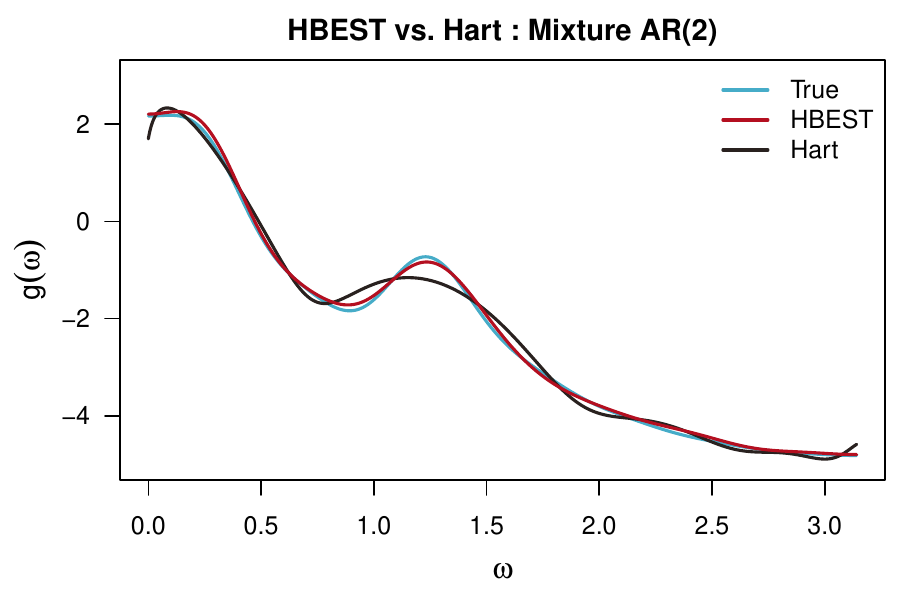}
    \includegraphics[width=.49\linewidth]{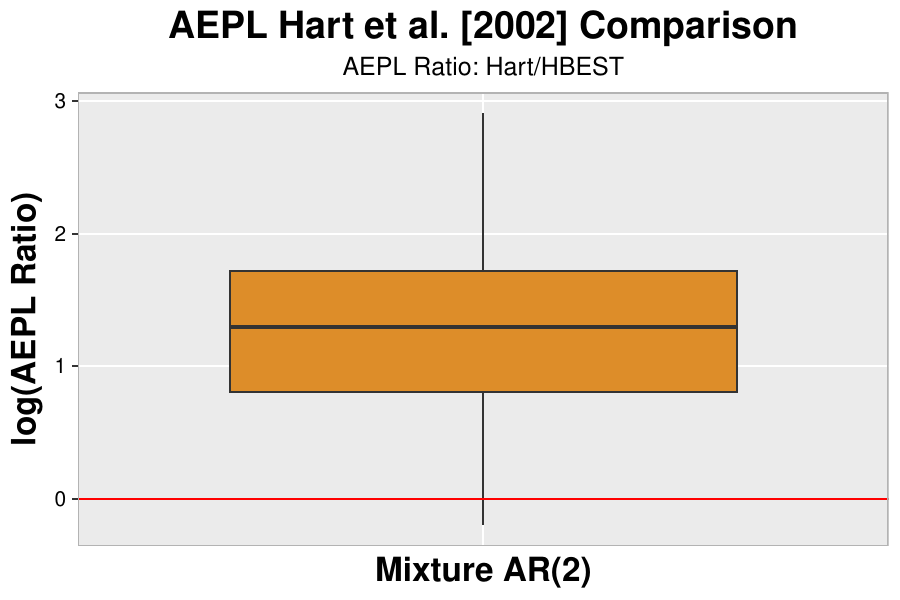}
    \caption{(left) Comparison of estimates for an individual log power spectrum using the proposed HBEST method and the method from \protect\cite{hart_nonparametric_2022}. (right) Distribution of log ratios of the Approximate Expected Posterior Losses (AEPLs) \eqref{eq.MSE_IndCom} across simulated datasets for HBEST and \protect\cite{hart_nonparametric_2022} for the conditional AR(2) mixture setting.
    Median and mean AEPLs for each model (HBEST, Hart) with lowest AEPL bolded: median: ($\pmb{0.03}$, $0.08$); mean: ($\pmb{0.03}$, $0.12$).} 
    \label{fig:Sims_mixture_comparison}
\end{figure}

Figure \ref{fig:Sims_mixture_comparison} shows examples of posterior mean estimates for the power spectrum of a single replicate for the conditional AR(2) mixture setting, as well as AEPL ratios. HBEST better captures the peaks in the log power spectrum leading to significantly improved estimation accuracy based on the AEPL ratios across datasets. 

In both cases, HBEST is significantly more computationally efficient than the Hart model with mean effective sample sizes per second of 0.836-0.909 for HBEST and 0.302-0.844 for the Hart model across all settings for estimating the population level log power spectrum.

\subsection{Comparison with Common and Independent Methods}\label{subsec.comparisonComInd}

Next, we compare our model against the two extreme competitors: the common method and the independent method. We simulated two levels of variability in $g_{\ell}$ across replicates $\ell$ to investigate the performance of HBEST under scenarios favorable to the common method (low variability) or the independent method (high variability).

\subsection{Hierarchical Data with Two Levels of Variation}\label{sec.generate_a_and_e_setting}

\begin{figure}[h!]
    \centering
    \includegraphics[width=.49\linewidth]{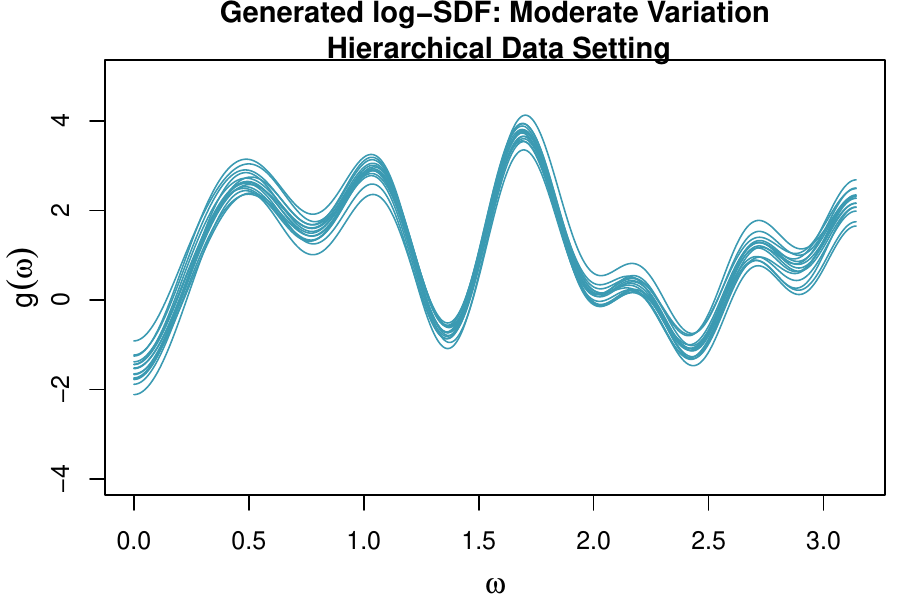}
    \includegraphics[width=.49\linewidth]{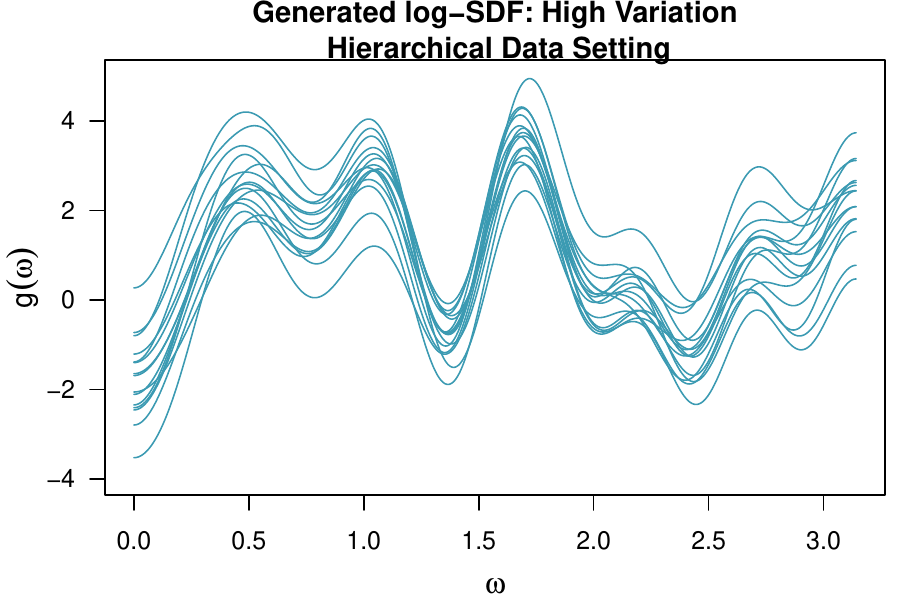}
    \caption{True individual level log power spectra for the two hierarchical data generation settings.} 
\label{fig:true_logSDF_IndCom_Comparison}
\end{figure}

In the hierarchical data setting, separately for each $S = 30$ data set, we generate $L = 15$ true $\alpha^*_{\ell} = \alpha^{\mathrm{*glob}} + \alpha^{\mathrm{*loc}}_{\ell}$, $\beta_{\ell, b}^* = \beta^{\mathrm{*glob}}_b + \beta^{\mathrm{*loc}}_{\ell, b}$ coefficients and then generate Gaussian time series $\{X_{\ell, t}\}$ with autocovariance functions given by $\gamma_{\ell}(h) = \int_{0}^{2\pi} \exp\{\alpha^*_{\ell}+\sum_{b=1}^{B}\psi_{b}(\omega) \beta_{\ell, b}^* + \imagine h\omega\} d\omega$. Specifically, we simulate the global scale parameter, $\tau^*$ from  $U(3,8)$, the local scale parameter $\zeta_{\ell}^*$ independently from $\mathcal{N}(x;\,0,1) \ind(1\leq x\leq 1.1)$, the global spline coefficients $\alpha^{\mathrm{*glob}}$ from $\mathcal{N}(0, 50/3)$, and $\beta^{\mathrm{*glob}}$ independently from $\mathcal{N}(0, \tau^{*2} d_b)$. We induce different levels of variation in the underlying spectral densities by setting scaling factor $\kappa > 0$ and then simulating local spline coefficients $\alpha^{\mathrm{*loc}}_{\ell}$ independently from $\mathcal{N}(0, \kappa \cdot 0.005)$ and $\beta^{\mathrm{*loc}}_{\ell, b}$ independently from $\mathcal{N}\left(0, \kappa \tau^{*2} d_b (\zeta^{*2}_{\ell}-1)\right)$. For the moderate variation setting, we set $\kappa = 0.1$, and for the high variation setting we set $\kappa = 1$. Finally, we generate the time series from $\mathcal{N}(\pmb{0}_{n_{\ell}}, \pmb{\Gamma}_{\ell})$, where $\pmb{\Gamma}_{\ell}$ is the $(n_{\ell} \times n_{\ell})$ Toeplitz autocovariance matrix formed from $\gamma_{\ell}(h)$, for $h = 0,\dots,n_{\ell} -1$. Figure \ref{fig:true_logSDF_IndCom_Comparison} shows examples of the generated true log power spectra from each setting.

To showcase that these models can innately handle differing length time series, we set   $n_{\ell} = 600$ for $80\%$ of the time series and $n_{\ell} = 1200$ for the remaining $20\%$. These lengths were chosen as they were close to the 1st and 3rd quartiles of the lengths from a sample of patients' first instance of non REM stage 3 sleep from the MESA data set. In the moderate variation setting, we set the small time series lengths to $n_{\ell} = 300$. We do this to explore how the HBEST model utilizes its inherent sharing of information under a setting where the common should perform well and the independent should suffer from smaller length time series.

\subsection{Results}\label{subsubsec.ResultsComInd}

\begin{figure}
    \centering
        \begin{subfigure}[b]{0.47\textwidth}
            \includegraphics[width=1\linewidth]{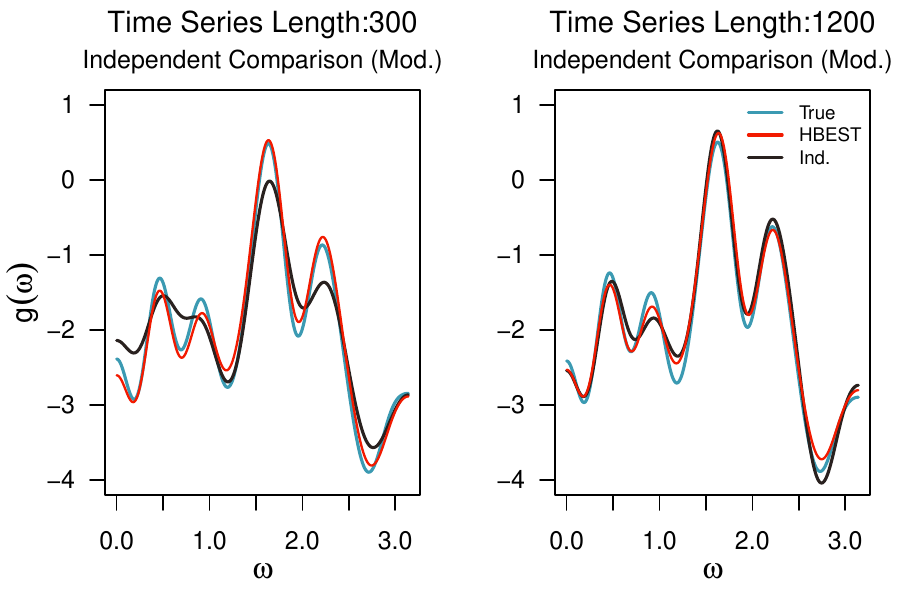}
            \caption{}
        \end{subfigure}
        \begin{subfigure}[b]{0.47\textwidth}
            \includegraphics[width=1\linewidth]{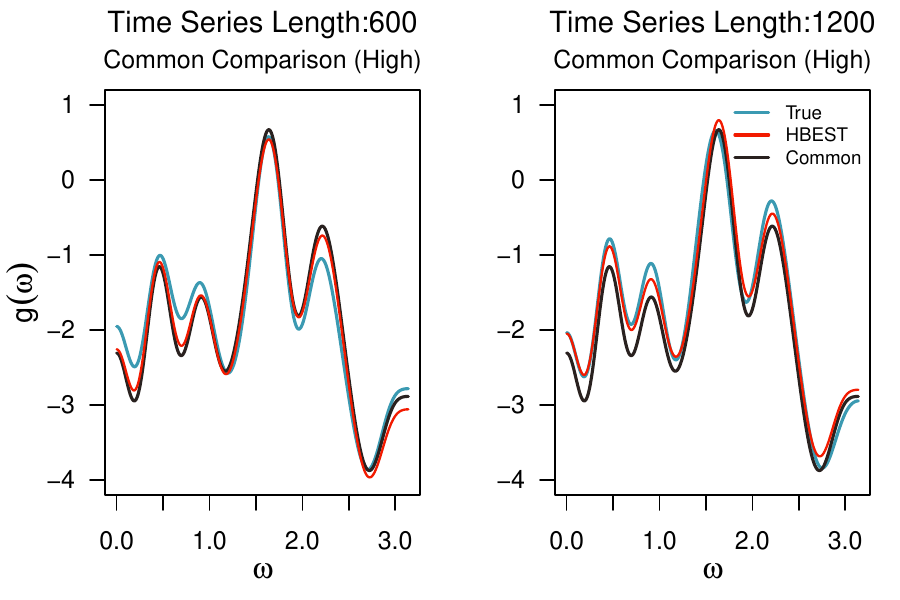} 
            \caption{}
        \end{subfigure}
        \caption{(a) Comparison of estimates for an individual log power spectra spectrum proposed HBEST method and the Independent method for the moderate variation hierarchical data generation setting. (b) Comparison of estimates for an individual log power spectrum proposed HBEST method and the Common method for the high variation hierarchical data generation setting.} 
\label{fig:Setting1_logSDF_comparison}
\end{figure}

\begin{figure}[h!]
    \centering
    \includegraphics[width=.6\linewidth]{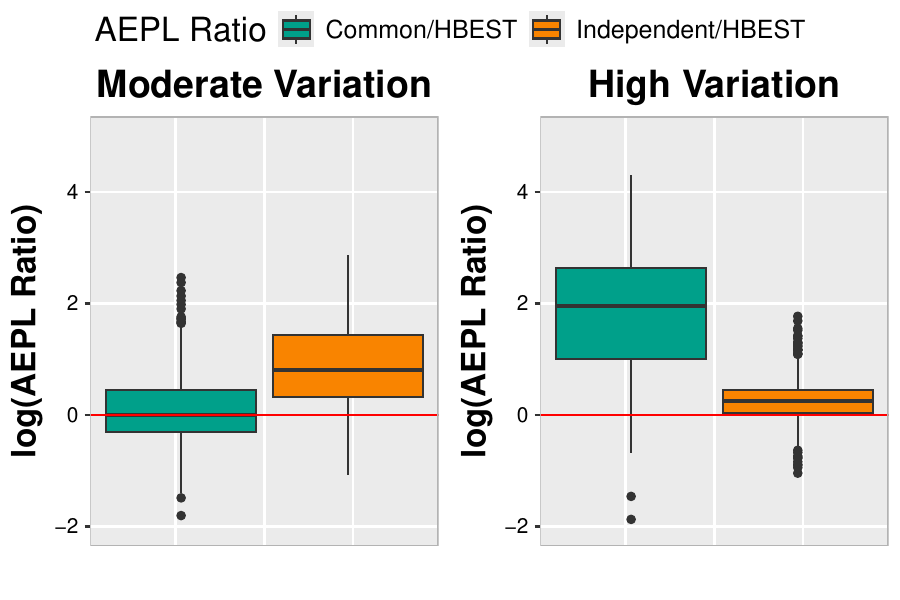}
    \caption{Hierarchical data setting: log of the AEPL ratio, where the AEPL is calculated with Equation \ref{eq.MSE_IndCom}. AEPLs for the median and mean for the moderate variation setting for each model (HBEST, Common, Independent) and lowest AEPL is bolded: median: ($\pmb{0.05}$, $0.06$, $0.16$); mean: ($\pmb{0.45}$, $0.48$, $0.55$). Statistics for the high variation setting: median: ($\pmb{0.06}$, $0.47$, $0.08$); mean: ($\pmb{0.39}$, $1.01$ ,$0.41$).} 
\label{fig:Setting1_logSDF_boxplot}
\end{figure}

Figure \ref{fig:Setting1_logSDF_comparison} shows the estimated individual log power spectra using the HBEST method and the competing methods for both settings. Figure \ref{fig:Setting1_logSDF_boxplot} visualizes the log of the \emph{non-trimmed} AEPL ratio calculations as defined in Equation \ref{eq.MSE_IndCom} and illustrates the competitiveness of the HBEST model against the Common and Independent methods. When the time series contain less variation across replicates, HBEST matches performance with the Common method and outperforms the Independent method. Likewise, when there is larger amounts of variation across the replicates, the HBEST model easily outperforms the Common method and modestly outperforms the Independent method. In both the moderate and high variation settings, HBEST had the lowest mean and median AEPL across all methods. 

\section{Heart Rate Variability and Cardiovascular Risk}
\label{s:app}
\begin{figure}[ht!]
 \centering
     \begin{subfigure}[b]{0.47\textwidth}
        \centering
    \includegraphics[width=1\linewidth]{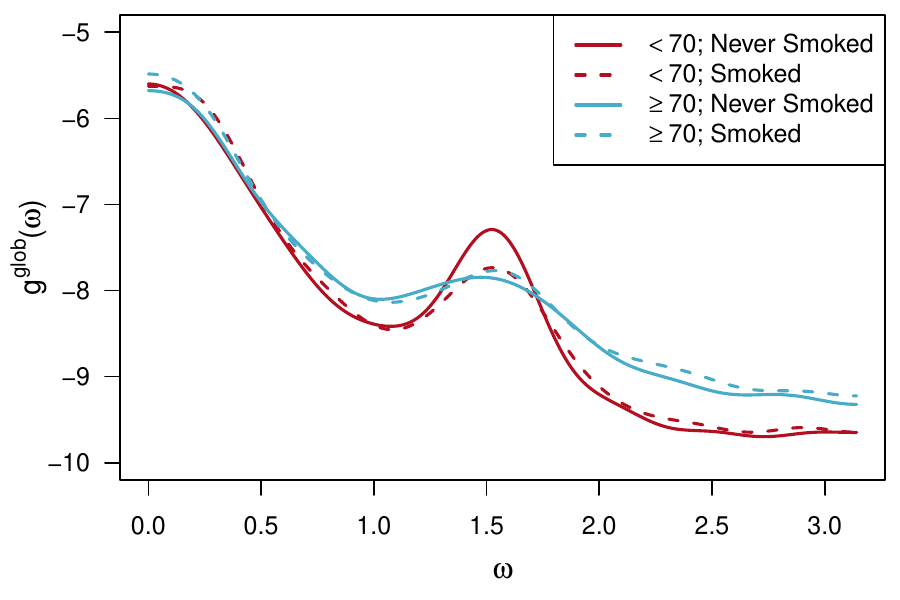}
      \caption{}
        \label{fig:mesa_global}
    \end{subfigure}
     \begin{subfigure}[b]{0.47\textwidth}
        \centering
    \includegraphics[width=1\linewidth]{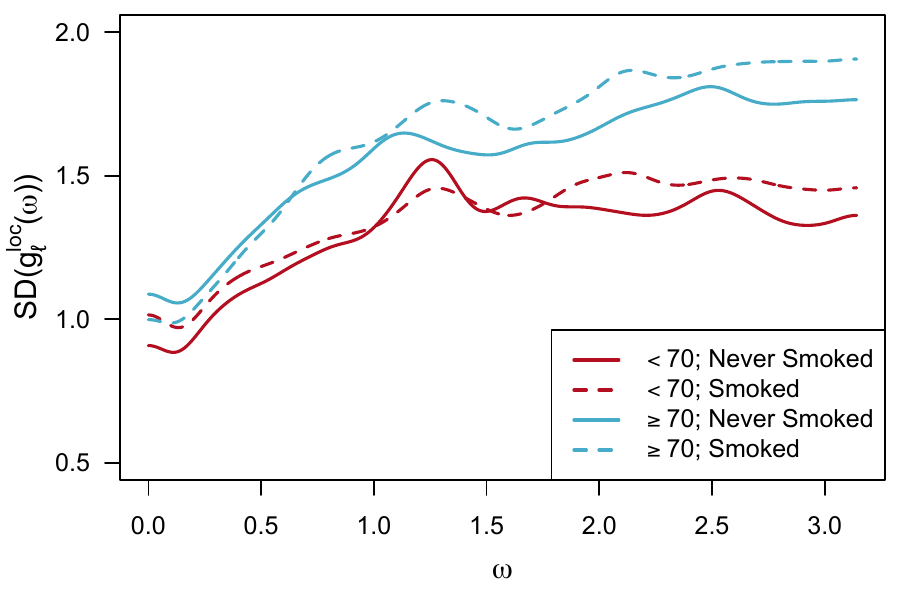}
      \caption{}
        \label{fig:mesa_sdlocal}
    \end{subfigure}
    
    \caption{(a) Posterior mean estimates of the subpopulation level log power spectra for 4 groups defined by smoking history (Never Smoked vs. Smoked) and age ($<$70 vs. $\ge 70)$.  (b) Standard deviation of the local estimates of the log power spectra within each subpopulation across frequencies.}
    \label{fig:MESA_allSDF}
\end{figure}

Returning to the motivating longitudinal study on cardiovascular disease (CVD) risk factors in a diverse population of older adults \citep{zhang_national_2018,chen_racialethnic_2015}, we now seek to characterize nocturnal HRV power spectra within this population and investigate associations with traditional cardiovascular disease risk factors,  age and smoking history.  Our analysis considers 1,151 MESA participants with a first onset of non-rapid eye movement sleep lasting at least 200 seconds and with relevant covariate information available. Since lower HRV high-frequency power is associated with an increased risk of cardiovascular disease \citep{hillebrandetal2013}, we focus our analysis on higher frequencies (0.15-0.4 Hz) \citep{HRVtaskforce96}. High frequency power reflects parasympathetic nervous system activity responsible for bodily activities that occur while at rest, and its modulation is inversely related to stress and arousal \citep{Shaffer2017HRV}. Figure \ref{fig:MESA_allSDF} presents HBEST estimates of subpopulation level log power spectra for 4 subgroups of MESA participants based on smoking history (Smoked vs. Never Smoked) and age ($<70$ vs. $\ge70$) and variability within these subpopulations.

These views provide insight into the association between high frequency HRV power during sleep and traditional risk factors for CVD. First, older adults tend to have lower high frequency power reflecting their relatively higher CVD risk relative to their younger counterparts.  Second, younger smokers have less high frequency power than younger non-smokers, which aligns with increased CVD risk associated with smoking.  Lastly, older adults exhibit more variability in the power spectra, especially among higher frequencies, which suggests that older adults exhibit more variability in CVD risk levels compared to younger adults.  Prior results from the scientific literature have also found that variability in CVD risk levels increases with age \citep{Tsai2020HealthyLifestyleCVD} thus aligning with the findings in this analysis. Taken together, these findings using HBEST align with previous studies \citep{hillebrandetal2013} suggesting that high frequency power in nocturnal HRV power spectra may serve as an important predictive biomarker that is indicative of CVD risk. 

\section*{Acknowledgments}

Research reported in this publication was supported by the National Institute Of General Medical Sciences (R01GM140476) and the National Science Foundation (CDS\&E-MSS-2152950). The content is solely the responsibility of the authors and does not necessarily represent the official views of the National Institutes of Health or the National Science Foundation. \\

\noindent The Multi-Ethnic Study of Atherosclerosis (MESA) Sleep Ancillary study was funded by NIH-NHLBI Association of Sleep Disorders with Cardiovascular Health Across Ethnic Groups (RO1 HL098433). MESA is supported by NHLBI funded contracts HHSN268201500003I, N01-HC-95159, N01-HC-95160, N01-HC-95161, N01-HC-95162, N01-HC-95163, N01-HC-95164, N01-HC-95165, N01-HC-95166, N01-HC-95167, N01-HC-95168 and N01-HC-95169 from the National Heart, Lung, and Blood Institute, and by cooperative agreements UL1-TR-000040, UL1-TR-001079, and UL1-TR-001420 funded by NCATS. The National Sleep Research Resource was supported by the National Heart, Lung, and Blood Institute (R24 HL114473, 75N92019R002).\vspace*{-8pt}

\clearpage
\bibliographystyle{apalike}
\bibliography{hbest_condensed}

\clearpage

\setcounter{section}{0}
\renewcommand{\thesection}{Appendix \Alph{section}}
\renewcommand{\thetable}{Table \Alph{table}}
\renewcommand{\thefigure}{Figure \Alph{figure}}
\def\theequation{A\arabic{equation}}

\section{Derivations}\label{app:derivations}

\subsection{Induced joint prior on spline coefficients}

The conditional priors placed on $\beta^{\mathrm{loc}}_{\ell b}$ and $\beta^{\mathrm{glob}}_b$ by HBEST are given by eq.~\eqref{eq:prior_betabl} in the main manuscript, respectively
\begin{align}\label{aeq:prior_betabl}
    \beta^{\mathrm{glob}}_b \mid \tau, d_b  \stackrel{\mathrm{ind}}{\sim} \calN(0, \tau^2 d_b), \qquad \beta^{\mathrm{loc}}_{\ell b} \mid \tau, d_b, \zeta_\ell \stackrel{\mathrm{ind}}{\sim} N\big(0, \tau^2 d_b (\zeta_\ell^2 - 1) \big).
\end{align}
The conditional element-wise independence and closure of the Gaussian family under convolution implies $\beta_{\ell b} \equiv \beta^{\mathrm{glob}}_b + \beta^{\mathrm{loc}}_{\ell b}$ is also Gaussian, conditioned on the hyperparameters. The expectation $\Ex(\beta_{\ell b}) = 0$ for all $\ell, b$; the covariance between $\beta_{\ell b}$ and $\beta_{\ell' b'}$ is
\begin{align*}
    \Cov(\beta_{\ell b}, \beta_{\ell' b'}) &= \Cov(\beta^{\mathrm{glob}}_b + \beta^{\mathrm{loc}}_{\ell b}, \beta^{\mathrm{glob}}_{b'} + \beta^{\mathrm{loc}}_{\ell' b'}) \\
    &= \Cov(\beta^{\mathrm{glob}}_b, \beta^{\mathrm{glob}}_{b'}) + \Cov(\beta^{\mathrm{glob}}_b, \beta^{\mathrm{loc}}_{\ell' b'}) + \Cov(\beta^{\mathrm{loc}}_{\ell b}, \beta^{\mathrm{glob}}_{b'}) + \Cov(\beta^{\mathrm{loc}}_{\ell b}, \beta^{\mathrm{loc}}_{\ell' b'}) \\
    &= \Var(\beta^{\mathrm{glob}}_b)\ind\{b=b'\} + 0 + 0 + \Var(\beta^{\mathrm{loc}}_{\ell b})\ind\{b = b', \ell = \ell'\} \\
    &= \tau^2 d_b \ind\{b = b'\} + \tau^2 d_b (\zeta_{\ell}^2 - 1) \ind\{b=b', \ell=\ell'\}.
\end{align*}
When $b \neq b'$, we have $\Cov(\beta_{\ell b}, \beta_{\ell, b'}) =0$; when $b = b, \: \ell \neq \ell'$, we have $\Cov(\beta_{\ell b}, \beta_{\ell, b'}) = \tau^2 d_b$; and when $b = b', \: \ell = \ell'$, we have $\Cov(\beta_{\ell b}, \beta_{\ell, b'}) = \tau^2d_b\zeta_{\ell}^2$. Correspondingly, let $\bB \equiv (\beta_{\ell b})$ denote the $\Re^{L \times B}$ matrix of spline coefficients, where $\bbeta_{\ell\bullet}$ is the $\ell^{\mathrm{th}}$ row of $\bB$ and $\bbeta_{\bullet b}$ is the $b^{\mathrm{th}}$ column of $\bB$. Entries within a row (where $b$ differs) retain joint independence, and entries within a column (where $\ell$ differs) have non-diagonal covariance, with the off-diagonal covariances constant at $\tau^2 d_b$. The induced conditional vector priors can thus be written
\begin{align*}
    \bbeta_{\ell\bullet} \mid \tau, \{d_b\}, \zeta_{\ell} &\sim \calN\left( \pmb{0}, \tau^2 (\zeta_{\ell}^2 - 1)\bD \right), \\
    \bbeta_{\bullet b} \mid \tau, d_b, \{\zeta_{\ell}\} &\sim \calN\left( \pmb{0}, \tau^2 d_b (\mathrm{diag}(\zeta_{\ell}^2 - 1) + \pmb{1}\pmb{1}^{\top}) \right).
\end{align*}
Finally, note that $\bW := \tau^2 \bD \otimes \left[ \mathrm{diag}(\zeta_{\ell}^2 - 1) + \pmb{1}\pmb{1}^{\top} \right] \in \Re^{LB \times LB}$ entry-wise consists of
\begin{align*}
    \bW_{Lb + \ell,Lb' + \ell'} &= \left[ \tau^2 \bD \right]_{b,b'} \left[ \mathrm{diag}(\zeta_{\ell}^2 - 1) + \pmb{1}\pmb{1}^{\top} \right]_{\ell,\ell'},
\end{align*}
where these evaluate to
\begin{align*}
    \begin{aligned}
        \bW_{Lb + \ell,Lb' + \ell'} &= 0 \times \left[ \mathrm{diag}(\zeta_{\ell}^2 - 1) + \pmb{1}\pmb{1}^{\top} \right]_{\ell,\ell'} = 0, &\qquad b &\neq b', \\
        \bW_{Lb + \ell,Lb' + \ell'} &= \tau^2 d_b \times 1 = \tau^2d_b, &\qquad b &= b', \: \ell \neq \ell', \\
        \bW_{Lb + \ell,Lb' + \ell'} &= \tau^2d_b\times [(\zeta_{\ell}^2 - 1) + 1] = \tau^2d_b\zeta^2_{\ell} , &\qquad b &= b', \: \ell = \ell'.
    \end{aligned}
\end{align*}
This matches the covariance structure of $\bB$, implying the conditional prior of $\bB$ is the matrix normal distribution given by
\begin{align*}
    \bB \mid \tau, \{d_b\}, \{\zeta_{\ell}\} &\sim \calM\calN\left( \pmb{0}, \mathrm{diag}(\zeta_{\ell}^2 - 1) + \pmb{1}\pmb{1}^{\top}, \tau^2 \bD \right)
\end{align*}

\clearpage
\section{MCMC Sampling Steps}

For computational simplicity we include the intercept terms, $\alpha^{\mathrm{glob}}$ and $\alpha^{\mathrm{loc}}_{\ell}$, as the respective zeroth entries in $\beta^{\mathrm{loc}}_{\ell b}$ and $\beta^{\mathrm{glob}}_b$, for $b = 0,1,\dots,B$ and $\ell = 1,\dots,L$. 
The sampling steps for each set of parameters $\{\beta^{\mathrm{loc}}_{\ell b}\}, \{\beta^{\mathrm{glob}}_b\}, \{\zeta_{\ell}\}$, and $\tau$ will be presented in vector notation, for which we adopt $\bbeta^{\mathrm{glob}} := (\begin{matrix} \beta_0^{\mathrm{glob}} & \cdots & \beta_B^{\mathrm{glob}} \end{matrix})^{\top}$ and $\bbeta^{\mathrm{loc}}_{\ell} := (\begin{matrix} \beta_{\ell 0}^{\mathrm{loc}} & \cdots & \beta_{\ell B}^{\mathrm{loc}} \end{matrix})^{\top}$; where necessary, we use $\bbeta_{-0}$ to denote $\bbeta$ absent the intercept term. We denote 
$\{\beta^{\mathrm{loc}}\} \equiv \{ \bbeta^{\mathrm{loc}}_{\ell} \}_{\ell=1}^{L}$, $\{\beta\} \equiv \{ \bbeta^{\mathrm{glob}}, \{\beta^{\mathrm{loc}}\}\}$, and $\{\zeta\} \equiv \{\zeta_{\ell}\}_{\ell=1}^L$. 
The corresponding priors are
\begin{align*}
    \begin{aligned}
        \bbeta^{\mathrm{glob}} \mid \tau &\sim \calN(0_{B+1}, \bSigma^{\mathrm{glob}}), &\enskip \bSigma^{\mathrm{glob}} &:= \left( \begin{matrix} \sigma^2_{\mathrm{glob}} & 0 \\ 0 & \tau^2 \bD \end{matrix} \right); \\
        \bbeta^{\mathrm{loc}}_{\ell} \mid \tau, \zeta_{\ell} &\stackrel{\mathrm{ind}}{\sim} \calN(0_{B+1}, \bSigma^{\mathrm{loc}}_{\ell}), &\enskip \bSigma^{\mathrm{loc}} &:= \left( \begin{matrix} \delta^2 & 0 \\ 0 & \tau^2 (\zeta_{\ell}^2 - 1)\bD \end{matrix} \right); \\
        \tau &\sim \mbox{Half-}t_{\nu_{\tau}}\ind(0, \infty); \\
        \zeta_{\ell} &\stackrel{\mathrm{iid}}{\sim} \mbox{Half-}t_{\nu_{\zeta}}\ind(1, \infty),
    \end{aligned}
\end{align*}
and the joint prior of our model is thus written
\begin{align}\label{eq.jointprior}
    \pi\left(\{\beta\}, \{\zeta\}, \tau\right) &= \left[ \prod_{\ell=1}^{L} \pi\left( \zeta_{\ell} \right) \pi(\bbeta^{\mathrm{loc}}_{\ell} \mid \tau, \zeta_{\ell}) \right] \pi\left(\bbeta^{\mathrm{glob}} \mid \tau \right) \pi(\tau).
\end{align}
The product Whittle likelihood utilizing the cosine basis expansion from eq.~\eqref{eq:DR_basic} is
\begin{equation}\label{eq.likelihood}
    \mathcal{L}\left( \{\beta\} \mid \{Y_{j_{\ell}}\} \right)  \propto \exp\left\{- \sum_{\ell=1}^{L} \pmb{1}^{\top}\bPsi_{\ell} ( \bbeta^{\mathrm{glob}} + \bbeta^{\mathrm{loc}}_{\ell} )  - \sum^{L}_{\ell=1} \sum_{j_{\ell}=1}^{m_{\ell}}\frac{Y_{j_{\ell}}}{\exp\left\{\bpsi_{j_{\ell}}^{\top} ( \bbeta^{\mathrm{glob}} + \bbeta^{\mathrm{loc}}_{\ell} )\right\}} \right\},
\end{equation}
where $m_{\ell} := \floor{ \frac{n_{\ell}}{2} }$, $\{Y_{j_{\ell}}\} \equiv \{Y_{j_{\ell}} : j_{\ell} = 1, \hdots, m_{\ell} \}_{\ell = 1}^L $ is the collection of replicate-specific periodograms, and $\bPsi_{\ell}$ is the matrix with rows $\bpsi_{j_{\ell}}^{\top} = ( \begin{matrix} 1 & \sqrt{2}\cos(1\omega_{j_{\ell}}) & \cdots & \sqrt{2}\cos(B\omega_{j_{\ell}}) \end{matrix} )$. Then the joint posterior distribution is 
\begin{equation}
    \pi(\{\beta\}, \{\zeta\}, \tau \mid Y)\propto\mathcal{L}( \{\beta\} \mid Y)\pi( \{\beta\}, \{\zeta\}, \tau).
\end{equation}
We use a Metropolis-within-Gibbs sampler to sequentially sample parameter values, first sampling the smoothing parameters and then the spline coefficient vectors.

\noindent\textbf{Sampling global smoothing parameter, $\tau$.} We sample $\tau$ with a ``Griddy Gibbs" sampler to avoid degeneracy at the boundary. Let $F_{\nu_{\tau}}$ denote the cumulative distribution function (CDF) for the $t$-distribution with degrees of freedom $\nu_{\tau}$ (i.e. of the prior for $\tau$), and let $F_{\nu_{\tau}}^{-1}$ denote the corresponding quantile function. For fixed bound parameters $0 < \tau_{\min} < \tau_{\max}$, we first construct a uniform grid $\{p_k\}$ of length $K$ between $F_{\nu_{\tau}}(\tau_{\min})$ and $F_{\nu_{\tau}}(\tau_{\max})$, and then construct a $\tau$-grid $\{\tau_{\mathrm{grid}}\} := \{t_k : t_k = F_{\nu_{\tau}}^{-1}(p_k)\}$. This final grid is essentially a discretization of the prior support for $\tau$; we sample from it with weights proportional to $\pi(\tau \mid -)$ to obtain a sample for $\tau$. This conditional posterior distribution is

\begin{align}
    \pi(\tau \mid -) &\propto \left[ \prod_{\ell=1}^L \pi(\bbeta^{\mathrm{loc}}_{\ell} \mid \tau, \zeta_{\ell}) \right] \pi(\bbeta^{\mathrm{glob}} \mid \tau) \pi(\tau), \nonumber\\
    \log \pi(\tau \mid -) &= -\frac{B(L + 1)}{2} \log (\tau^2) - \frac{1}{2\tau^2}\left( (\bbeta^{\mathrm{glob}}_{-0})^{\top}\bD^{-1}\bbeta^{\mathrm{glob}}_{-0} + \sum_{\ell=1}^L \frac{(\bbeta^{\mathrm{loc}}_{\ell,-0})^{\top}\bD^{-1}\bbeta^{\mathrm{loc}}_{\ell,-0} }{\zeta_{\ell}^2 - 1} \right) \nonumber\\
    &\phe \enskip - \frac{\nu_{\tau} + 1}{2} \log\left( 1 + \frac{\tau^2}{\nu_{\tau}} \right).\label{eq:logposterior_tau}
\end{align}
Then $\tau$ is sampled from the discrete distribution supported on $\{\tau_{\mathrm{grid}} \}$ with probabilities $\mathrm{P}[ \tau = t_k ] = \pi(t_k \mid -) / \sum_k \pi(t_k \mid -)$.

\noindent\textbf{Sampling local smoothing parameter, $\zeta_{\ell}$.}
For $\ell = 1, \dots, L$, we implement a Griddy Gibbs sampler to sample $\zeta_{\ell}$ from an approximated conditional posterior. Let $F_{\nu_{\zeta}}$ denote the CDF of the $t$-distribution with degrees of freedom $\nu_{\zeta}$, and let $F_{\nu_{\zeta}}^{-1}$ denote the corresponding quantile function. For fixed bound parameters $0 < \zeta_{\min} < \zeta_{\max}$, we first construct a uniform grid $\{p_k\}$ of length $K'$ between $F_{\nu_{\zeta}}(\zeta_{\min})$ and $F_{\nu_{\zeta}}(\zeta_{\max})$, and then construct a $\zeta$-grid $\{ \zeta_{\mathrm{grid}} \} := \{z_k : z_k = F_{\nu_{\zeta}}^{-1}(p_k)\}$. We sample from this discrete support with weights proportional to the conditional posterior distribution $\pi(\zeta_{\ell} \mid -)$ to obtain a sample for $\zeta_{\ell}$. This conditional posterior distribution is
\begin{align}
    \pi(\zeta_{\ell} \mid -) &\propto \pi(\bbeta^{\mathrm{loc}}_{\ell} \mid \tau, \zeta_{\ell}) \pi(\zeta_{\ell}), \nonumber\\
    \log \pi(\zeta_{\ell} \mid -) &\propto -\frac{B}{2} \log (\zeta_{\ell}^2 - 1) - \frac{(\bbeta^{\mathrm{loc}}_{\ell,-0})^{\top}\bD^{-1}\bbeta^{\mathrm{loc}}_{\ell,-0}}{2\tau^2 (\zeta_{\ell}^2 - 1)}  -\frac{\nu_{\zeta} + 1}{2} \log \left( 1 + \frac{\zeta^2}{\nu_{\zeta}} \right).\label{eq:logposterior_zeta}
\end{align}
Then $\zeta_{\ell}$ is sampled from the discrete distribution supported on $\{\zeta_{\mathrm{grid}} \}$ with probabilities $\mathrm{P}[ \zeta_{\ell} = z_k ] = \pi(z_k \mid -) / \sum_k \pi(z_k \mid -)$.

\noindent\textbf{Sampling local spline vector, $\bbeta^{\mathrm{loc}}_{\ell}$.} 
For $\ell = 1, \cdots, L$, we propose a new local spline coefficient vector with a scaled-Laplace approximation to the conditional posterior distribution for $\bbeta^{\mathrm{loc}}_{\ell}$. We numerically obtain the conditional posterior mode $\widetilde{\bbeta}{}^{\mathrm{loc}}_{\ell}$ through gradient ascent, and use the Hessian $\bH(\widetilde{\bbeta}{}^{\mathrm{loc}}_{\ell})$ to determine the covariance of the Gaussian approximation. The conditional posterior distribution is proportionally dependent on the prior's Gaussian kernel and the Whittle likelihood by
\begin{align}\label{eq:local_logpost}
    \pi( \bbeta^{\mathrm{loc}}_{\ell} \mid -) &\propto \exp\left\{ - \pmb{1}^{\top}\bPsi_{\ell} \bbeta^{\mathrm{loc}}_{\ell} - \sum_{j_{\ell}=1}^{m_{\ell}}\frac{Y_{j_{\ell}}}{\exp\left\{\bpsi_{j_{\ell}}^{\top}\left( \bbeta^{\mathrm{glob}} + \bbeta^{\mathrm{loc}}_{\ell} \right)\right\}}  - \frac{1}{2}(\bbeta^{\mathrm{loc}}_{\ell})^{\top} (\bSigma^{\mathrm{loc}}_{\ell})^{-1}\bbeta^{\mathrm{loc}}_{\ell} \right\}; \nonumber\\
    \log \pi(\bbeta^{\mathrm{loc}}_{\ell} \mid -) &= - \pmb{1}^{\top}\bPsi_{\ell} \bbeta^{\mathrm{loc}}_{\ell} - \sum_{j_{\ell}=1}^{m_{\ell}}\frac{Y_{j_{\ell}}}{\exp\left\{\bpsi_{j_{\ell}}^{\top}\left( \bbeta^{\mathrm{glob}} + \bbeta^{\mathrm{loc}}_{\ell} \right)\right\}}  - \frac{1}{2}(\bbeta^{\mathrm{loc}}_{\ell})^{\top} (\bSigma^{\mathrm{loc}}_{\ell})^{-1}\bbeta^{\mathrm{loc}}_{\ell},
\end{align}
where we omit the normalizing constant. For brevity, let $\lambda_{j_{\ell}} := Y_{j_{\ell}} \exp\{ -\bpsi_{j_{\ell}}^{\top}(\bbeta^{\mathrm{glob}} + \bbeta^{\mathrm{loc}}_{\ell}) \}$ denote the summand entries, and let $\bLambda_{\ell} := \mathrm{diag}(\begin{matrix} \lambda_1 & \cdots & \lambda_{m_{\ell}} \end{matrix})$.
The gradient and Hessian with respect to $\bbeta^{\mathrm{loc}}_{\ell}$ are
\begin{align} 
    \nabla \log \pi(\bbeta^{\mathrm{loc}}_{\ell} \mid -) &= -\bPsi_{\ell}^{\top}\pmb{1} - (\bSigma^{\mathrm{loc}}_{\ell})^{-1} \bbeta^{\mathrm{loc}}_{\ell} + \sum_{j_{\ell}=1}^{m_{\ell}} \bpsi_{j_{\ell}} \lambda_{j_{\ell}}; \label{eq:local_gradient}\\
    \bH(\bbeta^{\mathrm{loc}}_{\ell}) := \nabla^2 \log \pi(\bbeta^{\mathrm{loc}}_{\ell} \mid -) &= -(\bSigma^{\mathrm{loc}}_{\ell})^{-1} - \sum_{j_{\ell}=1}^{m_{\ell}} \bpsi_{j_{\ell}} \bpsi_{j_{\ell}}^{\top} \lambda_{j_{\ell}} = -(\bSigma^{\mathrm{loc}}_{\ell})^{-1} - \bPsi_{\ell}^{\top} \bLambda_{\ell} \bPsi_{\ell}. \label{eq:local_hessian}
\end{align}
Note the Hessian is negative definite. We find $\widetilde{\bbeta}{}^{\mathrm{loc}}_{\ell} := \argmax \{ \log \pi(\bbeta^{\mathrm{loc}}_{\ell} \mid -) \}$ numerically using \textsf{R}'s \texttt{stats::optim} function, and sample proposal value $\bbeta^{\mathrm{loc}}_* \sim \calN(\widetilde{\bbeta}{}^{\mathrm{loc}}_{\ell}, -\eta\bH(\widetilde{\bbeta}{}^{\mathrm{loc}}_{\ell})^{-1})$, where $\eta > 0$ is a fixed scalar selected to improve mixing; typically $\eta = 1$ suffices. We then accept $\bbeta^{\mathrm{loc}}_{\ell} \leftarrow \bbeta^{\mathrm{loc}}_*$ (or not) with a Metropolis-Hastings step.

\noindent\textbf{Sampling global spline vector, $\bbeta^{\mathrm{glob}}$.} We also sample $\bbeta^{\mathrm{glob}}$ via Laplace approximation. The conditional log posterior of $\bbeta^{\mathrm{glob}}$ is
\begin{align}\label{eq:global_logpost}
    \log \pi(\bbeta^{\mathrm{glob}} \mid -) &= -\sum_{\ell=1}^L \pmb{1}^{\top}\bPsi_{\ell} \bbeta^{\mathrm{glob}} - \sum_{\ell=1}^L \sum_{j_{\ell}=1}^{m_{\ell}} \frac{Y_{j_{\ell}}}{\exp\{ \bpsi_{j_{\ell}}^{\top}\bbeta^{\mathrm{glob}} \} \exp\{ \bpsi_{j_{\ell}}^{\top}\bbeta^{\mathrm{loc}}_{\ell} \}},
\end{align}
where again the normalizing constant has been omitted. Using the condensed notation of $\lambda_{j_{\ell}}$ and $\bLambda_{\ell}$ for each $\ell = 1, \hdots, L$, the gradient and Hessian of the conditional log posterior with respect to $\bbeta^{\mathrm{glob}}$ is
\begin{align}
    \nabla \log \pi(\bbeta^{\mathrm{glob}} \mid -) &= -\sum_{\ell=1}^L \bPsi_{\ell}^{\top}\pmb{1} - (\bSigma^{\mathrm{glob}})^{-1} \bbeta^{\mathrm{glob}} + \sum_{\ell=1}^L \sum_{j_{\ell}=1}^{m_{\ell}} \bpsi_{j_{\ell}} \lambda_{j_{\ell}} ; \label{eq:global_gradient}\\
    \bH(\bbeta^{\mathrm{glob}}) := \nabla^2 \log \pi(\bbeta^{\mathrm{glob}} \mid -) &= -(\bSigma^{\mathrm{glob}})^{-1} - \sum_{\ell=1}^{L} \bPsi_{\ell}^{\top} \bLambda_{\ell} \bPsi_{\ell}. \label{eq:global_hessian}
\end{align}
We find $\widetilde{\bbeta}{}^{\mathrm{glob}} := \argmax \{ \log \pi(\bbeta^{\mathrm{glob}} \mid -) \}$ numerically using \textsf{R}'s \texttt{stats::optim} function, and sample proposal value $\bbeta^{\mathrm{glob}}_* \sim \calN(\widetilde{\bbeta}{}^{\mathrm{glob}}, -\eta \bH( \widetilde{\bbeta}{}^{\mathrm{glob}} )^{-1})$. We then accept $\bbeta^{\mathrm{glob}} \leftarrow \bbeta^{\mathrm{glob}}_*$ (or not) with a Metropolis-Hastings step.

\clearpage
\section{MCMC Algorithms}

\begin{algorithm}[!h]
\caption{MCMC Algorithm: Sampling Scheme}
\label{alg::MCMCSteps}
\begin{algorithmic}[1]
\setstretch{1.25}
\State Initialize ($i=1$) parameters $\tau$, $\{ \zeta_{\ell} \}$, $\{ \beta^{\mathrm{loc}}_{\ell} \}$, $\bbeta^{\mathrm{glob}}$
\For{$i \in \{2, \hdots, I\}$} \Comment{MCMC iterations}
    \State \textbf{Step 1: Griddy Gibbs for $\tau$:}
    \Indent
    \State Sample $\tau$ using a Griddy Gibbs step (see Algorithm \ref{alg:griddygibbs_tau}).
    \State Update $\bSigma^{\mathrm{glob}} \gets \mathrm{diag}( \begin{matrix} \sigma_{\mathrm{glob}}^2 & \tau^2 d_1 & \cdots & \tau^2 d_B \end{matrix})$.
    \EndIndent
    \State \textbf{Step 2: Griddy Gibbs for each $\zeta_{\ell}$:}
    \Indent
        \For{$\ell \in \{1, \dots, L\}$}
            \State Sample $\zeta_{\ell}$ using a Griddy Gibbs step (see Algorithm \ref{alg:griddygibbs_zeta}).
            \State Update $\bSigma^{\mathrm{loc}} \gets \mathrm{diag}( \begin{matrix} \delta^2 & \tau^2 (\zeta_{\ell}^2 - 1) d_1 & \cdots & \tau^2 (\zeta_{\ell}^2 - 1)d_B \end{matrix})$.
        \EndFor
    \EndIndent
    \State \textbf{Step 3: Metropolis-Hastings for each $\bbeta^{\mathrm{loc}}_{\ell}$:}
    \Indent
        \For{$\ell \in \{1, \dots, L\}$}
            \State Sample proposal $\bbeta^{\mathrm{*loc}}_{\ell}$ using a Laplace approximation.
            \State Accept/reject $\bbeta^{\mathrm{*loc}}_{\ell}$ with a Metropolis-Hastings step (see Algorithm \ref{alg:laplace_local}).
        \EndFor
    \EndIndent
    
    \State \textbf{Step 4: Metropolis-Hastings for $\bbeta^{\mathrm{glob}}$:}
    \Indent
        \State Sample proposal $\bbeta^{\mathrm{*glob}}$ using a Laplace approximation.
        
        \State Accept/reject $\bbeta^{\mathrm{*glob}}$ with a Metropolis-Hastings step (see Algorithm \ref{alg:laplace_global}).
    \EndIndent
    \State \textbf{Step 5: Store samples.}
\EndFor
\vspace{4pt}
\end{algorithmic}
\end{algorithm}

\clearpage
\begin{algorithm}[!h]
\caption{Griddy Gibbs Sampler for $\tau$}
\label{alg:griddygibbs_tau}
\begin{algorithmic}[1]
\setstretch{1.25}
\Procedure{SampleGlobalScale}{$\{\beta\}, \{\zeta_{\ell}\}, \bD, \nu_\tau > 0, 0 < \tau_{\min} < \tau_{\max}, K \in \mathbb{N}$}
    \State $p_{\min} \gets F_{\nu_{\tau}}(\tau_{\min})$.
    
    \State $p_{\max} \gets F_{\nu_{\tau}}(\tau_{\max})$.
    
    \State $\{p_{\mathrm{grid}}\} \gets \{p_k = p_{\min} + (p_{\max} - p_{\min}) \frac{k - 1}{K - 1} \::\: k = 1, \dots, K\}$.
    
    \State $\{ \tau_{\mathrm{grid}} \} \gets \{ t_k = F_{\nu_{\tau}}^{-1}(p_k) \::\: p_k \in \{p_{\mathrm{grid}}\} \}$.
    
    \For{$k \in \{1, \hdots, K\}$}
        \State Compute $\log \pi(t_k \mid -)$ using eq.~\eqref{eq:logposterior_tau}.
    \EndFor
    
    \For{$k \in \{1, \hdots, K\}$}
        \State Compute $\log \widetilde{\pi}_k \gets \log \pi(t_k \mid -) - \max_j \{\log \pi(t_j \mid -)\}$.
    \EndFor
    
    \For{$k \in \{1, \hdots, K\}$}
        \State Compute $p_k^* \gets \exp \{ \log \widetilde{\pi}_k \} / \sum_j \exp \{ \log \widetilde{\pi}_j \}$ \: (i.e. \texttt{softmax}).
    \EndFor
    
    \State Sample $\tau \sim \text{Discrete}(\{\tau_{\mathrm{grid}}\}, \{p_k^*\})$.
    
    \State \Return $\tau$.
    \vspace{4pt}
\EndProcedure
\end{algorithmic}
\end{algorithm}

\clearpage
\begin{algorithm}[!h]
\caption{Griddy Gibbs Sampler for $\zeta_{\ell}$}
\label{alg:griddygibbs_zeta}
\begin{algorithmic}[1]
\setstretch{1.25}
\Procedure{SampleLocalScale}{$\bbeta^{\mathrm{glob}}, \bbeta^{\mathrm{loc}}_{\ell}, \tau, \bD, \nu_{\zeta} > 0, 0 < \zeta_{\min} < \zeta_{\max}, K' \in \mathbb{N}$}
    \State $p_{\min} \gets F_{\nu_{\zeta}}(\zeta_{\min}).$
    
    \State $p_{\max} \gets F_{\nu_{\zeta}}(\zeta_{\max}).$
    
    \State $\{p_{\mathrm{grid}}\} \gets \{p_k = p_{\min} + (p_{\max} - p_{\min}) \frac{k - 1}{K' - 1} \::\: k = 1, \dots, K'\}$.
    
    \State $\{ \zeta_{\mathrm{grid}} \} \gets \{ z_k = F_{\nu_{\zeta}}^{-1}(p_k) \::\: p_k \in \{p_{\mathrm{grid}}\} \}$.
    
    \For{$k \in \{1, \hdots, K'\}$}
        \State Compute $\log \pi(z_k \mid -)$ using eq.~\eqref{eq:logposterior_zeta}.
    \EndFor
    
    \For{$k \in \{1, \hdots, K'\}$}
        \State Compute $\log \widetilde{\pi}_k \gets \log \pi(z_k) - \max_j \{\log \pi(z_j)\}$.
    \EndFor
    
    \For{$k \in \{1, \hdots, K'\}$}
        \State Compute $p_k^* \gets \exp \{ \log \widetilde{\pi}_k \} / \sum_j \exp \{ \log \widetilde{\pi}_j \}$ \: (i.e. \texttt{softmax}).
    \EndFor
    
    \State Sample $\zeta \sim \text{Discrete}(\{\zeta_{\mathrm{grid}}\}, \{p_k^*\})$.
    
    \State \Return $\zeta$.
    \vspace{4pt}
\EndProcedure
\end{algorithmic}
\end{algorithm}

\clearpage
\begin{algorithm}[!h]
\caption{Laplace Approximation Metropolis-Hastings Sampler for $\bbeta^{\mathrm{loc}}_{\ell}$}
\label{alg:laplace_local}
\begin{algorithmic}[1]
\setstretch{1.25}
\Procedure{SampleLocalSpline}{$\{\beta\}, \tau, \zeta_{\ell}, Y, \bD, \{\bPsi_{\ell}\}_{\ell=1}^L, \eta > 0$}

    \State Compute MAP estimate $\widetilde{\bbeta}{}^{\mathrm{loc}}_{\ell} \gets \argmax \log \pi(\bbeta^{\mathrm{loc}}_{\ell} \mid -)$; see eq.~\eqref{eq:local_logpost}–\eqref{eq:local_hessian}.

    \State Compute Hessian $\bH(\widetilde{\bbeta}{}^{\mathrm{loc}}_{\ell}) \gets \nabla^2 \log \pi(\widetilde{\bbeta}{}^{\mathrm{loc}}_{\ell} \mid -)$; see eq.~\eqref{eq:local_hessian}.

    \State Sample proposal $\bbeta^{\mathrm{*}} \sim \calN(\widetilde{\bbeta}{}^{\mathrm{loc}}_{\ell}, - \eta \cdot \bH(\widetilde{\bbeta}{}^{\mathrm{loc}}_{\ell})^{-1})$.

    \State Compute acceptance ratio
    \begin{align*}
        \alpha^* \gets \min \left\{ 1, \: \frac{\pi(\bbeta^* \mid -)}{\pi(\bbeta^{\mathrm{loc}}_{\ell,\mathrm{curr.}} \mid -)} \right\}; \qquad \mbox{see eq.~\eqref{eq:local_logpost}.}
    \end{align*}

    \State Sample $U \sim \mathrm{unif}(0, 1)$.
    
    \If{$U < \alpha^*$}
        \State Accept proposal $\bbeta^{\mathrm{loc}}_{\ell} \gets \bbeta^*$.
    \Else
        \State Reject proposal $\bbeta^{\mathrm{loc}}_{\ell} \gets \bbeta^{\mathrm{loc}}_{\ell,\mathrm{curr.}}$.
    \EndIf
    \vspace{4pt}
\EndProcedure
\end{algorithmic}
\end{algorithm}

\clearpage
\begin{algorithm}[!h]
\caption{Laplace Approximation Metropolis-Hastings Sampler for $\bbeta^{\mathrm{glob}}$}
\label{alg:laplace_global}
\begin{algorithmic}[1]
\setstretch{1.25}
\Procedure{SampleGlobalSpline}{$\{\beta\}, \tau, Y, \bD, \{\bPsi_{\ell}\}_{\ell=1}^L, \eta > 0$}

    \State Compute MAP estimate $\widetilde{\bbeta}{}^{\mathrm{glob}} \gets \argmax \log \pi(\bbeta^{\mathrm{glob}} \mid -)$; see eq.~\eqref{eq:global_logpost}–\eqref{eq:global_hessian}.

    \State Compute Hessian $\bH(\widetilde{\bbeta}{}^{\mathrm{glob}}) \gets \nabla^2 \log \pi(\widetilde{\bbeta}{}^{\mathrm{glob}} \mid -)$; see eq.~\eqref{eq:global_hessian}.

    \State Sample proposal $\bbeta^{\mathrm{*}} \sim \calN(\widetilde{\bbeta}{}^{\mathrm{glob}}, - \eta \cdot \bH(\widetilde{\bbeta}{}^{\mathrm{glob}})^{-1})$.

    \State Compute acceptance ratio
    \begin{align*}
        \alpha^* \gets \min \left\{ 1, \: \frac{\pi(\bbeta^* \mid -)}{\pi(\bbeta^{\mathrm{glob}}_{\mathrm{curr.}} \mid -)} \right\}; \qquad \mbox{see eq.~\eqref{eq:global_logpost}.}
    \end{align*}

    \State Sample $U \sim \mathrm{unif}(0, 1)$.
    
    \If{$U < \alpha^*$}
        \State Accept proposal $\bbeta^{\mathrm{glob}} \gets \bbeta^*$.
    \Else
        \State Reject proposal $\bbeta^{\mathrm{glob}} \gets \bbeta^{\mathrm{glob}}_{\mathrm{curr.}}$.
    \EndIf
    \vspace{4pt}
\EndProcedure
\end{algorithmic}
\end{algorithm}

\end{document}